\documentclass[traditabstract]{aa} 
\usepackage[english]{babel}	
\usepackage{graphicx}
\usepackage{txfonts}
\usepackage{natbib}
\usepackage[bookmarks=false]{hyperref}
\usepackage{verbatim}
\usepackage{longtable, booktabs}
\usepackage{float}

\def\hi{\ifmmode{\rm HI}\else{H\/{\sc i}}\fi} 
\newcommand {\kms} {\,{\rm km\,s}^{-1}}

\newcommand {\kpc} {\,{\rm kpc}}

\newcommand {\Gyr} {{\rm Gyr}}
\newcommand {\de}{^{\circ}}
\newcommand {\mo}{{\rm M}_\odot}

\newcommand {\moyr}{\,{\rm M_\odot\,\rm yr}^{-1}}

\begin{document}

  \title{Gas accretion from minor mergers in local spiral galaxies}

  \subtitle{}

  \author{E. M. Di Teodoro
         \inst{1}
	  \and
	  	F. Fraternali
	  	\inst{1,2}
      }
 \authorrunning{Di Teodoro \& Fraternali}
 \institute{
           Department of Physics and Astronomy, University of Bologna, 6/2, Viale Berti Pichat, 40127 Bologna, Italy
           \and
           Kapteyn Astronomical Institute, Postbus 800, 9700 AV Groningen, The Netherlands
         }

  \date{}

\abstract{
In this paper we quantify the gas accretion rate from minor mergers onto star-forming galaxies in the Local Universe using \hi\ observations of 148 nearby spiral galaxies (WHISP sample). We developed a dedicated code that iteratively analyses \hi\ data-cubes, finds dwarf gas-rich satellites around larger galaxies and estimates an upper limit to the gas accretion rate. We found that 22\% of the galaxies have at least one detected dwarf companion. We made the very stringent assumption that all satellites are going to merge in the shortest possible time transferring all their gas to the main galaxies. This leads to an estimate of the maximum gas accretion rate of 0.28 $\moyr$, about five times lower than the average SFR of the sample. Given the assumptions, our accretion rate is clearly an overestimate. Our result strongly suggests that minor mergers do not play a significant role in the total gas accretion budget in local galaxies.}


\maketitle

\section{Introduction}

The evolution of galaxies is strongly affected by their capability of retaining their gas and accreting fresh material from the surrounding environment. Galaxies belonging to the so-called  ``blue-sequence'', which are actively forming stars and are dominated by young stellar populations, show an almost constant or a slowly declining star formation rate (SFR) throughout the Hubble time \citep[e.g.,][]{Panter+07}. Since the gas consumption time-scales are always of the order of a few Gyrs \citep{Noeske+07, Bigiel+11}, spiral galaxies need to replenish their gas at rates comparable to their star formation rates \citep{Hopkins+08, Fraternali&Tomassetti12}. These arguments are fully applicable to the Milky Way: with a SFR of 1-3 $\mo$ slowly declining over the last $\sim10$ Gyrs \citep[e.g.,][]{Aumer&Binney09, Chomiuk&Povich11}, the Galaxy would have exhausted its gas reservoir in a few Gyrs without replacement from outside \citep[e.g.,][]{Chiappini+97}. 

There are essentially two sources from which disc galaxies can gain new gas: the intergalactic medium (IGM) and other gas-rich galaxies. The IGM is the place where the most of baryons are thought to still reside \citep[e.g.,][]{Bregman07}. Most of this gas should be in a diffuse warm-hot phase \citep[e.g.,][]{Shull+12}. Therefore the IGM represents a huge reservoir of nearly pristine gas but how this material can cool and accrete onto the discs is not well understood.  Current cosmological simulations predict that gas accretion can occur in two modes \citep[e.g.,][]{Ocvirk+08, Keres+09}: the ``hot'' accretion, which dominates the growth of massive galaxies, and the ``cold'' accretion through filamentary streams and clouds, which prevails in lower mass structures and at high redshifts \citep[e.g.,][]{Dekel&Birnboim06}. 

The second channel for gas accretion is given by merger events. According to the Extended Press-Schechter theory, the structures in the Universe grow by several inflowing events and have increased their mass content through a small number of major mergers, more common at high redshifts, and through an almost continuous infall of dwarf galaxies \citep{Bond+91,Lacey&Cole93}. Although several theoretical \citep[e.g.,][]{Stewart+09, Kazantzidis+09} and observational studies \citep[e.g.,][]{Patton+00, Lotz+08, Lambas+12} have been carried out in the last years, the predictions and the estimates for the galaxy merger rate and its evolution with redshift remain uncertain and no consensus has been achieved yet \citep[e.g.,][]{Bertone&Conselice09, Hopkins+10}. 

In this paper, we use neutral hydrogen (\hi) observations to investigate gas accretion from minor mergers onto star-forming galaxies in the Local Universe. The advantage of using \hi\ observations instead of the optical-UV ones is that both morphological and kinematical information are immediately available. In addition, the gas layers are more easily disturbed by tidal interactions than the stellar disc. Two recent studies, namely \cite{Holwerda+11} and \cite{Sancisi+08}, have taken advantage of \hi\ data and both make use of the WHISP catalogue \citep{vanderHulst+01}.  \cite{Holwerda+11} focused on the galaxy merger fraction and, employing techniques developed for optical-UV observations, found a merger fraction between 7\% and 13\%. Instead, \cite{Sancisi+08} attempted to quantify  the contribution of minor mergers to the total gas accretion. They found that 25\% of local galaxies show signs of minor interactions or have disturbed \hi\ distribution and, assuming lifetimes for these observed features of about 1 Gyr and typical accreted \hi\ mass of order $10^8$-$10^9\, \mo$, they calculated an accretion rate of about 0.1-0.2 $\moyr$. This value is about an order of magnitude lower than typical star formation rates. 

In this study, we use a quantitative approach to obtain a reliable estimate for the merger fraction and the gas accretion rate. We make use of a specific-purpose numerical code that is able to quickly analyse a large number of \hi\ data-cubes, find dwarf gas-rich companions around disc galaxies and estimate an upper limit for the gas accretion onto the discs. In section \ref{sec:method}, we describe the main features of our code and the methods adopted. In section \ref{sec:application}, we show the results obtained by applying our analysis on a sub-sample of the WHISP catalogue and we discuss them in section \ref{sec:discussion}.

\section{Method}\label{sec:method}

We wrote a numerical code to automatically identify 3D sources in data-cubes, i.e. image arrays with two spatial dimensions, related to the position on the plane of the sky, and one spectral dimension, which can correspond either to velocity or to frequency. Our code is targeted to work with \hi\ observations as it performs a three dimensional scanning of the data to look for dwarf gas-rich companions around large galaxies. Once the program has found a candidate, it derives its physical properties, such as the \hi\ mass, the projected distance from the main galaxy and an estimate of the accretion rate onto the central disc. In short, the code used in this work is essentially a source finder plus an algorithm for estimating the accretion rate. 

The standard flow of our code can be outlined in three steps:

\begin{enumerate}
\item \emph{Identifying the main-galaxy}. The pixels referable to the central galaxy emission are identified and isolated through  an appropriate mask.

\item \emph{Searching for dwarf galaxies}. The data-cube is scanned for three-dimensional sources and dwarf galaxies or \hi\ clouds inside the field of view are identified. 

\item \emph{Estimating the gas accretion rate}. For each detected dwarf, a maximum accretion rate onto the disc is calculated by estimating a minimum time of collision between the satellite and the central galaxy.
\end{enumerate}

In the following sections, we describe the most important steps and the main features and limitations of our code.

\subsection{Searching for sources and background statistics}
The searching algorithm is derived from \emph{Duchamp}, a code dedicated to three-dimensional source detection \citep{Whiting12} and developed for the Australian SKA Pathfinder (ASKAP). The basic idea behind this algorithm is to locate and connect groups of bright and contiguous pixels that lie above some flux threshold, without imposing any size or shape requirement to the detection. The search is performed using either a two-dimensional raster-scanning algorithm \citep{Lutz80} or a one-dimensional research along each individual pixel spectrum. Three-dimensional sources (two spatial dimensions and one spectral) are then built up on the basis of adjacency or neighbourhood criteria both in the velocity and in the spatial domain. For a full description of the source finding algorithm, we refer to the \emph{Duchamp} main paper. The one-dimensional technique is less computationally expensive, but it can bring to spurious detections. On the contrary, the Lutz algorithm is generally more reliable at the price of the computational slowness. As discussed later in this section, we used the 1-D method to identify the main galaxy emission and the 2-D technique to detect companions.

The searching algorithm can be schematically summarized as follows:

\begin{enumerate}
\item \emph{Pixel detection}. The data-cube is scanned using one of the  above mentioned techniques and a list of all pixels with a flux greater than a given threshold is produced.

\item \emph{Merging objects}. The detected pixels that are considered close to each other based on spatial and spectral requirements are merged together. Adjacent detections or detections lying within a user-defined range of pixels or channels are expected to belong to the same object. After this step, a list of three-dimensional sources is produced.

\item \emph{Growing objects}. The size of the detections is increased by adding pixels at the edges of the objects that are above some secondary threshold. This step guarantees a smooth transition between the source and the background.

\item \emph{Rejecting objects}. Not all detected objects can be considered reliable. Sources that do not match some agreement criteria, like a minimum number of contiguous pixels and channels, are discarded.
\end{enumerate}

A crucial point of the searching algorithm is the determination of the flux threshold above which a pixel can be considered as a part of a source. In order to do this, the code needs to estimate the central value and width of the noise distribution in the data-cube. The former should be zero or very close to zero for \hi\ data-cubes without systematics (due for instance to problems with the data reduction). The typical data-cube of the WHISP survey is dominated by a large number of noise pixels and a relatively small number of bright pixels that belong to the sources. In such a situation, it is preferable to calculate the noise over the whole data-cube using the median as noise middle and the median absolute deviation from the median (MADFM, hereinafter) as noise spread. These quantities are less sensitive to the presence of very bright pixels than the equivalent normal statistics, the mean $\mu$ and the standard deviation $\sigma$. For a Gaussian distribution, the standard deviation can be written in term of the MADFM as $\sigma=s/0.6745$ \citep[for details, see][and references therein]{Whiting12}. The threshold $\tau$ is then determined with a simple sigma-clipping, i.e., it is set at a number $n$ of standard deviations $\sigma$ above the median $m$:

\begin{equation}
\label{eq:threshold}
 \tau = m + n\sigma
 \end{equation}

Such a value is the minimum flux that a pixel must possess to be selected by the algorithm. We checked that the noise middle and spread calculated using the whole data-cube are the same as those obtained using only the channels with line emission; the differences do not exceed 5\%.

We used the searching algorithms in two different steps: the identification of the main galaxy and the detection of satellites. The former consists in isolating all those regions ascribable to the main galaxy emission. The code performs a search in the data-cube using the one-dimensional technique and selects as the main galaxy the object that covers the largest number of pixels. This approach is not computationally expensive, and it is reliable when the code is analysing an heterogeneous group of galaxies, but it does not allow the code to identify systems in advanced phase of merging, i.e., when a companion is physically connected in space and velocity with the main galaxy. Concerning the satellites, we used the Lutz algorithm, which guarantees a better reliability and minimizes the number of spurious detections. We impose a neighbourhood criterion grounded on the spatial and spectral resolution of the observations: each detected pixel is merged with other detected pixels lying within a spatial beam and within two velocity channels, which is the typical instrumental broadening (FWHM) for \hi\ observations when Hanning smoothing has been applied. Finally, we reject all those detections that are smaller than the beam area of the observations and less extended in velocity than the spectral broadening. 
We stress that a three-dimensional source finding algorithm, unlike the two-dimensional methods, can isolate sources with different kinematics even if they are totally or partially overlapped in the plane of the sky. Indeed, if two sources have radial velocities that differ more than the typical velocity resolution ($\sim$10-15 $\kms$), they are always detected as separate sources, no matter whether they overlap or not in the sky.

\subsection{Accretion and star formation rate estimates}

The main purpose of this study is to estimate the maximum gas accretion rate coming from minor mergers. In the following we describe our assumptions. 

\begin{figure}[t]
\centering
\includegraphics[width=0.5\textwidth]{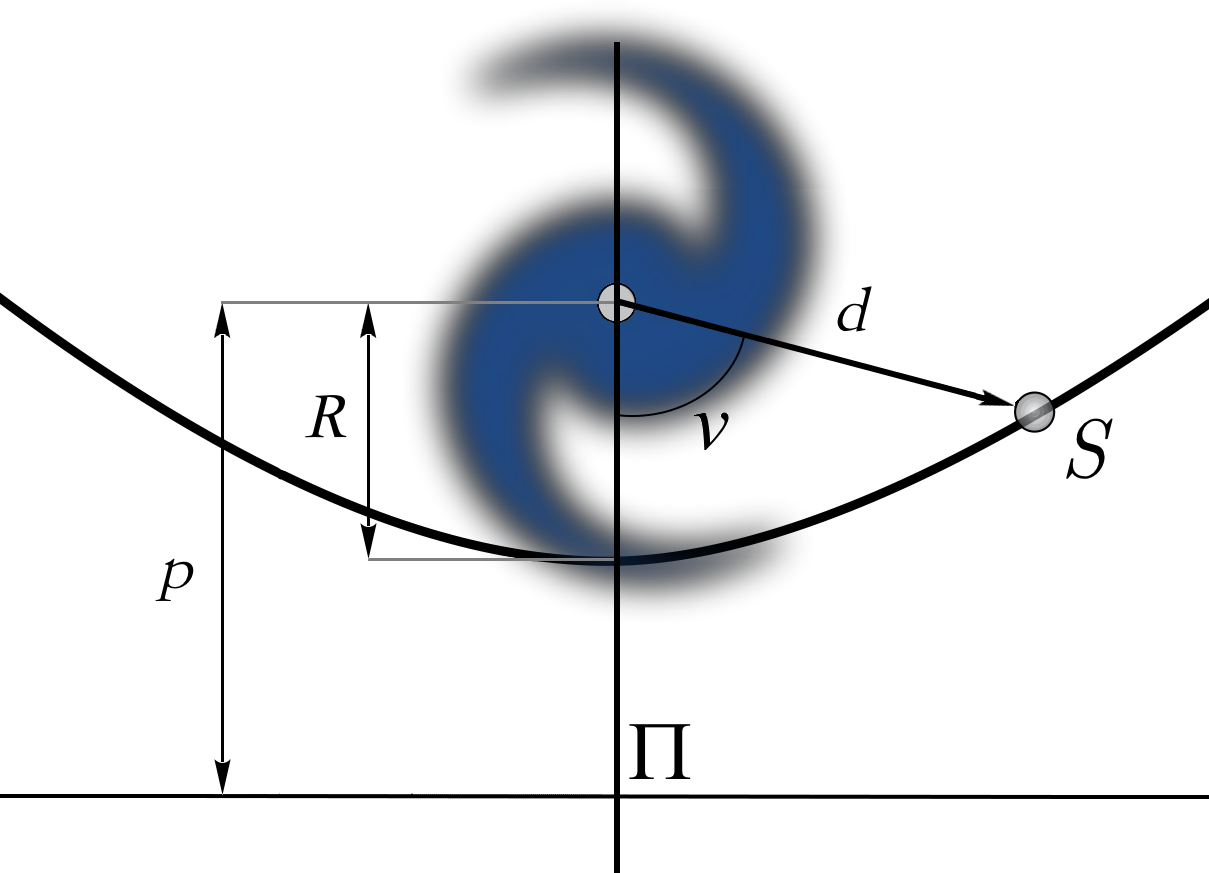}
\caption{Schematical view of the parabolic orbit approximation. The blue spiral is the main galaxy, \emph{S}  is the satellite with projected distance \emph{d} and true anomaly \emph{$\nu$}. The distance \emph{p} between the center of the spiral galaxy and the directrix $\Pi$ of the parabola is two times the outer radius of the main galaxy \emph{R}. }
\label{fig:parabola}
\end{figure}

Firstly, we assume that all dwarf galaxies will collide in the future with the main galaxies and that their gas will be entirely and instantaneously accreted. Secondly, we assume that the collision will occur in the shortest possible time. In order to calculate this time, we make the satellites moving in parabolic trajectories leading to impact the outer regions of the main galaxies. The orbit is defined in the three-dimensional space by fixing the focus of the parabola at the centre of the main galaxy, imposing the passage through the satellite and fixing the position of the orbital peri-centre at a distance equal to the maximum radius of the central galaxy (Fig. \ref{fig:parabola}). For a generic conic orbit, the time-scale of collision can be obtained by using the equation of the true anomaly $\nu$ of celestial mechanics:

\begin{equation}\label{eq:true0}
\int_0^\nu {\frac{d\nu'}{(1-e\cos\nu')^2}} = \sqrt{\frac{\mu}{p^3}}\;(t-T_0)
\end{equation}

\noindent where $e$ is the eccentricity of the orbit, $T_0$ is the time of the peri-centre passage, $p$ is the semi-latus rectum of the conic section and $\mu = G(M_\mathrm{main} + M_\mathrm{sat}) \sim GM_\mathrm{main}$ is the total dynamical mass of the system galaxy plus satellite multiplied by the gravitational constant G. The dynamical mass of the central galaxy $M_\mathrm{main}(R_\mathrm{max})=G^{-1}v_\mathrm{c}(R_\mathrm{max})^2R_\mathrm{max}$ is calculated within the maximum radius $R_\mathrm{max}$ of the source, estimated by the searching algorithm. The circular velocity $v_\mathrm{c}$ is obtained from the velocity widths of the \hi\ global profiles at the 20\% of the peak flux corrected for the inclination taken from the HyperLEDA catalogue.
Solving the integral (\ref{eq:true0}) for parabolic orbits ($e = 1$) one obtains the following formula which describes the variation of the true anomaly $\nu$ as a function of time:

\begin{equation}
\label{eq:true}
\tan\frac{\nu}{2}+\frac{1}{3}\tan^3\frac{\nu}{2} = \sqrt{\frac{\mu}{p^3}}\;(t-T_0)
\end{equation}

\noindent where the semi-latus rectum for parabolic orbit is $p = 2R$, being $R$ the distance between the focus and the vertex of the parabola (Fig. \ref{fig:parabola}).

Using equation (\ref{eq:true}), we can estimate the time of the peri-centre passage $T_0$ by calculating $\nu$ through a de-projection of the projected anomaly $\nu_\mathrm{p}$ of the dwarf galaxy measured in the plane of the sky. The accretion rate of cold gas onto a certain galaxy is then obtained by dividing the \hi\ mass of each dwarf by its time of peri-centre passage:

\begin{equation}
\label{eq:rate}
\dot{M}_\mathrm{HI} = \sum_{i=0}^n M_{\mathrm{HI}\,,\,i}/T_{0\,,\,i}
\end{equation}

\noindent where the sum is taken over all the detected companion galaxies. The \hi\ mass $M_\mathrm{HI}$ is calculated from the flux density using the following relation \citep{Roberts75}:

\begin{equation}
\label{eq:HImass}
M_\mathrm{HI} = 2.356 \times 10^5 \, D^2 \, \int S(v) \, dv
\end{equation}

\noindent where $\int S(v) \, dv$ is the integral across the line of the flux density corrected for the primary beam attenuation and expressed in Jy km s$^{-1}$ and $D$ is the distance in Mpc. The equation (\ref{eq:HImass}) is valid under the assumption that the gas is optically thin, which is generally a good approximation for neutral hydrogen, especially in dwarf galaxies, thus no correction for \hi\ self-absorption was applied. The distances were preferably taken from the Extragalactic Distance Database \cite[EDD,][]{Tully+09}, available for a number of galaxies with $v_\mathrm{sys}$ up to $10000 \, \kms$ and mostly obtained from Cepheids, TRGB, SNIa or Cosmicflows-2 project \citep{Tully+13}. Otherwise, we used the NASA/IPAC Extragalactic Database (NED). For seven galaxies with no available better estimates, we used the Hubble flow $D=v_\mathrm{sys}/H_0$ with $H_0=70$ km s$^{-1}$ Mpc$^{-1}$ and the systemic velocity $ v_\mathrm{sys}$  corrected for Virgo-centric inflow using the values given by the HyperLEDA catalogue.

We compare the total gas accretion (\ref{eq:rate}) to the gas depletion due to the star formation process in the discs. The SFR of the central galaxies was calculated from the far-infrared luminosities \citep{Kennicutt98}:

\begin{equation}
\label{eq:kenn}
\mathrm{SFR} = \frac{L_\mathrm{FIR}}{2.2 \times 10^{43}} \quad \mathrm{M_\odot \, yr^{-1} }
\end{equation}

\noindent with the $L_\mathrm{FIR}$ in erg s$^{-1}$ obtained from the far-infrared flux FIR defined after \cite{Helou+85} as:

\begin{equation}
\label{eq:FIR}
\mathrm{FIR} = 1.26\times10^{-11} (2.58f_{60\mu} + f_{100\mu}) \quad \mathrm{erg \, s^{-1} \, cm^{-2}} 
\end{equation}

\noindent where $f_{60\mu}$ and $f_{100\mu}$ are the fluxes at 60 and 100 micron expressed in Jansky. In this work we used the IRAS fluxes taken from NED and HyperLEDA. All main galaxies in our sample are detected both at $60\mu$ and $100\mu$. See Tab. \ref{tab:maingalaxies} for their main physical properties.

\subsection{Major and minor mergers}\label{baryonicmass}

We split major and minor mergers depending on the baryonic mass ratio: pair of galaxies with $M_\mathrm{bar, sat}/M_\mathrm{bar, main}\leq0.20$ are classified as minor mergers, otherwise as major. We preferably estimate the baryonic mass as:

\begin{equation}
\label{eq:baryonicmass}
M_\mathrm{bar} = M_\mathrm{*} + 1.4M_\mathrm{HI}
\end{equation}

\noindent where the factor 1.4 take into account the helium gas fraction. We neglected the contribution of molecular gas. The \hi\ mass $M_\mathrm{HI}$ is directly estimated from data-cubes through eq. (\ref{eq:HImass}). A rough estimate of the stellar mass $M_\mathrm{*}$ is obtained by using the total K$_\mathrm{s}$-band magnitude, corrected for extinction, taken from the 2MASS Redshift Survey \cite[2MRS,][]{Huchra+12} and adopting the following formula \citep[e.g.][]{Longhetti&Saracco09}:

\begin{equation}
\label{eq:stellarmass}
\log_{10}(M_\mathrm{*}) = \log_{10}(M/L_\mathrm{K}) -0.4[K+5-5\log_{10}(D_\mathrm{[pc]}) -3.28]
\end{equation}

\noindent where $M/L_\mathrm{K}$ is the stellar mass-to-light ratio (in solar units) in the K-band and 3.28 is the absolute K-band magnitude of the Sun in the Vega system \citep{Binney&Merrifield98}. We assumed a constant value of mass-to-light ratio $M/L_\mathrm{K}=0.6\, \mo/L_{\odot,\mathrm{K}}$, compatible with stellar population models \citep[e.g.][]{Portinari+04} with a Kroupa IMF \citep{Kroupa02}. 
 
When 2MRS magnitudes were not available, namely for most dwarf satellites and a few main galaxies, we directly derived $M_\mathrm{bar}$ from the Baryonic Tully-Fisher Relation (BTFR): 

\begin{equation}
\label{eq:BTF}
\log_{10} (M_\mathrm{bar})= a\log_{10} (v_\mathrm{flat})+b
\end{equation}

\noindent with $a=3.82\pm0.22$ and $b = 2.01 \pm 0.41$  \citep{McGaugh12}. The $v_\mathrm{flat}$ was assumed as half of the inclination-corrected velocity widths $w_{20}$ of the \hi\ global profiles at the 20\% of the peak flux. Since inclination angles are not known for most dwarf satellites, we adopted an average inclination of 60 degrees for these galaxies.

\subsection{The data sample}
The Westerbork \hi\ survey of Irregular and Spiral galaxies Project \citep[WHISP,][]{vanderHulst+01} is a survey of the neutral hydrogen content in galaxies selected from the Uppsala General Catalogue \citep[UGC,][]{Nilson73} and observed with the Westerbork Synthesis Radio Telescope (WSRT). WHISP is to date the largest publicly available catalogue of \hi\ nearby galaxies observed with an interferometer and it includes galaxies at $\delta > 20\de$ (B1950) with major axis apparent size $>$ 1.2$'$ (B band) and \hi\ flux densities $F_\mathrm{HI} > 100$ mJy. Objects satisfying these selection criteria have generally systemic velocities less than 6000 $\kms$, i.e., distances lower than 85 Mpc using the Hubble flow with $H_0=70$ km s$^{-1}$ Mpc$^{-1}$. The galaxies were chosen to be reasonably distributed over all Hubble types, even if later-type galaxies are favoured by the observational criteria. The highest spatial resolution for the WHISP data is 12$''$ x 12$''$/sin $(\delta)$, the typical channel separation is of the order of 5 $\kms$. In this work, we used both \hi\ data-cubes spatially smoothed to 30$''$ x 30$''$ and 60$''$ x 60$''$. The original sample comprises 256 data-cubes containing about 370 galaxies.\footnote{ The datacubes, the column density maps and the velocity fields of the WHISP galaxies, at 12$''$, 30$''$ and 60$''$ of resolution, are publicly available for the ``Westerbork on the Web'' project at ASTRON (http://www.astron.nl/wow/).}

\begin{figure}[t]
\centering
\includegraphics[width=0.5\textwidth]{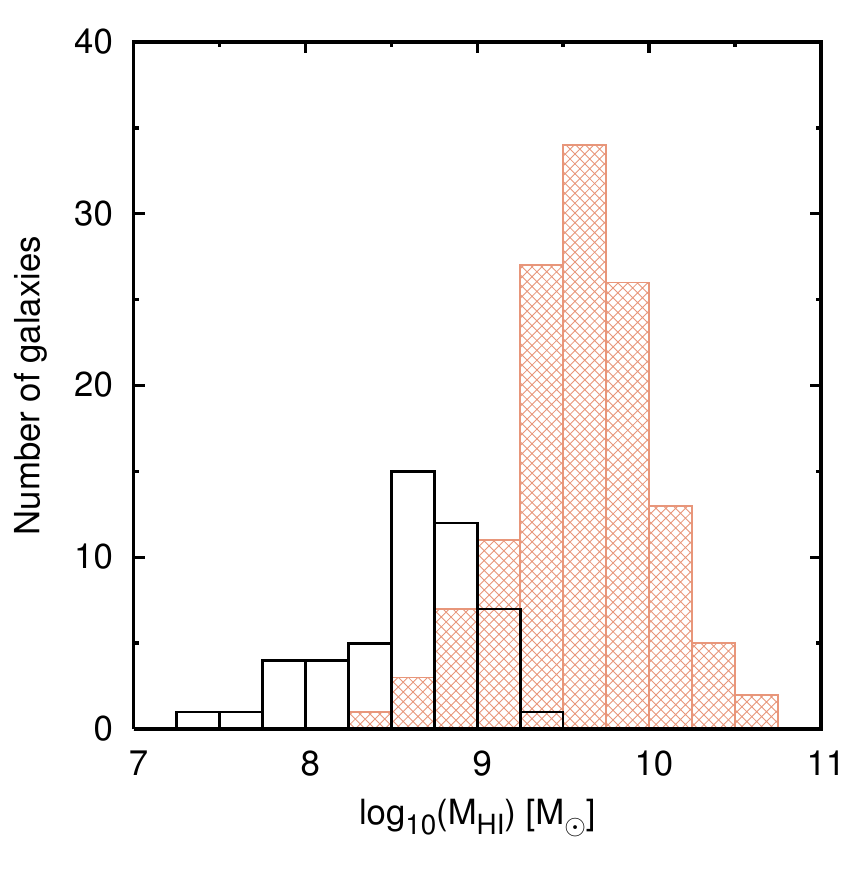}
\caption{\hi\ Mass distribution of the detected galaxies in the WHISP sample. Orange shadowed boxes show the spiral galaxies selected as v$_\mathrm{flat}>100 \, \kms$, black boxes show their dwarf satellites. }
\label{fig:MF}
\end{figure}

Since our goal is to study dwarf satellites around large star-forming galaxies, we selected a sub-sample of spiral galaxies by keeping only  those data-cubes containing at least one galaxy with rotation velocity $v_\mathrm{flat}= w_{20}/(2\sin i)>100 \, \kms$. The selection was performed through a cross-correlation between the $w_{20}$ estimated directly from the data-cubes and the $w_{20}$ calculated using the Tully-Fisher relation from \cite{Sakai+00}:

\begin{equation}\label{eq:TF}
M_B = -(7.97\pm0.72)(\log w_{20}-2.5)-(19.80\pm0.11)
\end{equation}

\noindent where $M_B$ is the B-band absolute magnitude (corrected for galactic extinction and k-correction), taken from HyperLEDA. We kept only galaxies for which both methods returned $v_\mathrm{flat}>100 \, \kms$. This cross-correlation is needed to avoid spurious selections related to some unreliable inclination angles in the HyperLEDA catalogue. Our final sample has 148 data-cubes. Spiral galaxies therein have usually neutral hydrogen masses between $10^9 \; \mo$ and few $10^{10} \; \mo$ (Fig. \ref{fig:MF}). The global properties of the main galaxies are listed in Tab. \ref{tab:maingalaxies}.

\section{Results}\label{sec:application}

We ran our code both on data-cubes smoothed to $30''$ and $60''$. The results obtained with these two data sets are thoroughly comparable. We fixed a sigma-clipping threshold for the source finder equal to 4 (see equation \ref{eq:threshold}) and a secondary threshold for growing objects at the edges of 2.5. After extensive experiments, these values appeared the best compromise between reaching low sensitivities and avoiding spurious detections.

\begin{figure*}[!p]
\centering
\includegraphics[width=0.95\textwidth]{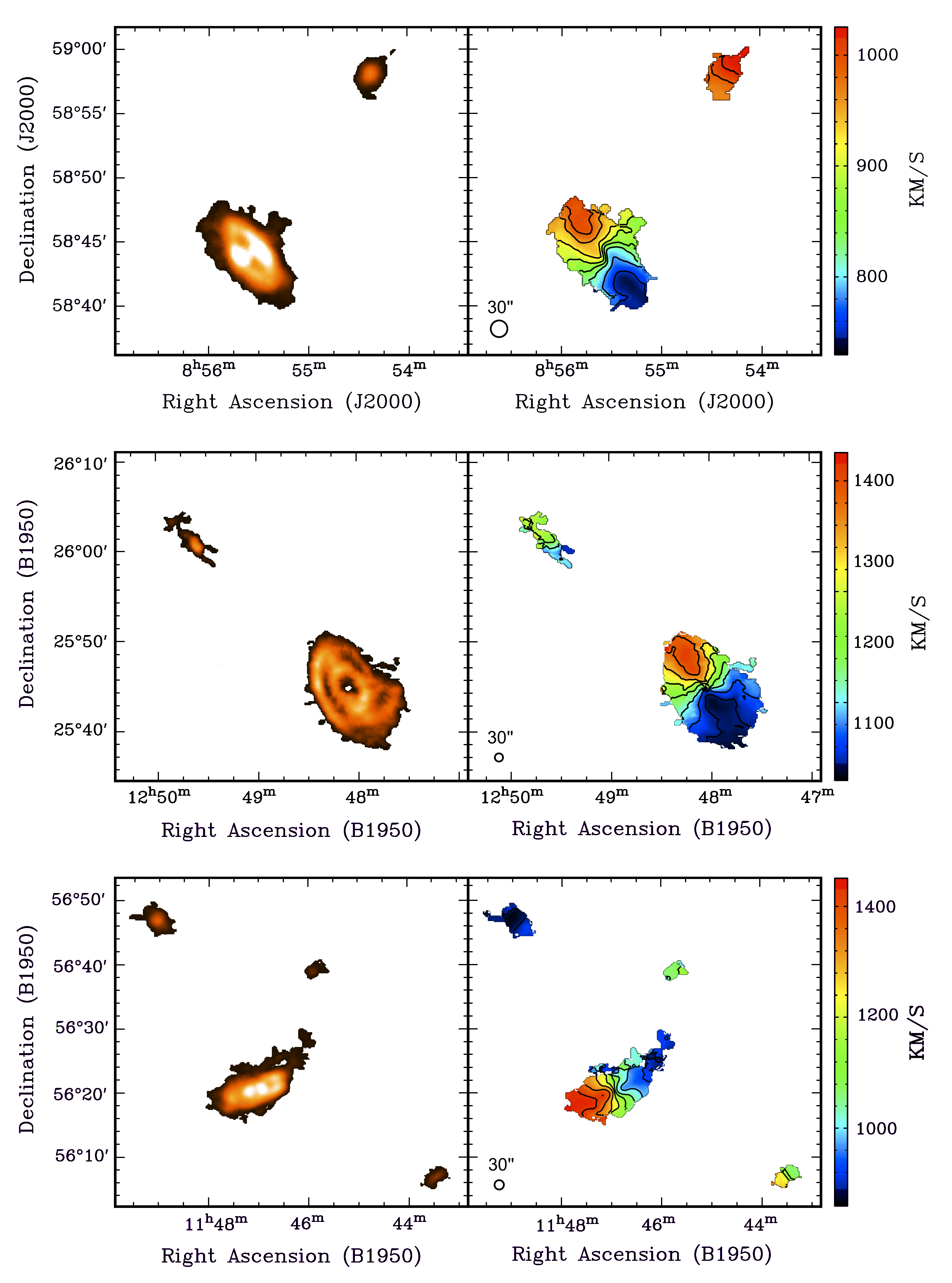}
\caption{ Three examples of multiple systems in the WHISP sample. From the \emph{top} to the \emph{bottom}, UGC 4666, UGC 7989, UGC 6787 and their dwarf companions. In the \emph{left panels}, the \hi\ column-density maps (0th moment), in the \emph{right panels}, the velocity fields (1st moment) obtained from 30'' smoothed data-cubes.  }
\label{fig:examples}
\end{figure*}

\begin{figure*}[!t]
\centering
\includegraphics[width=0.99\textwidth]{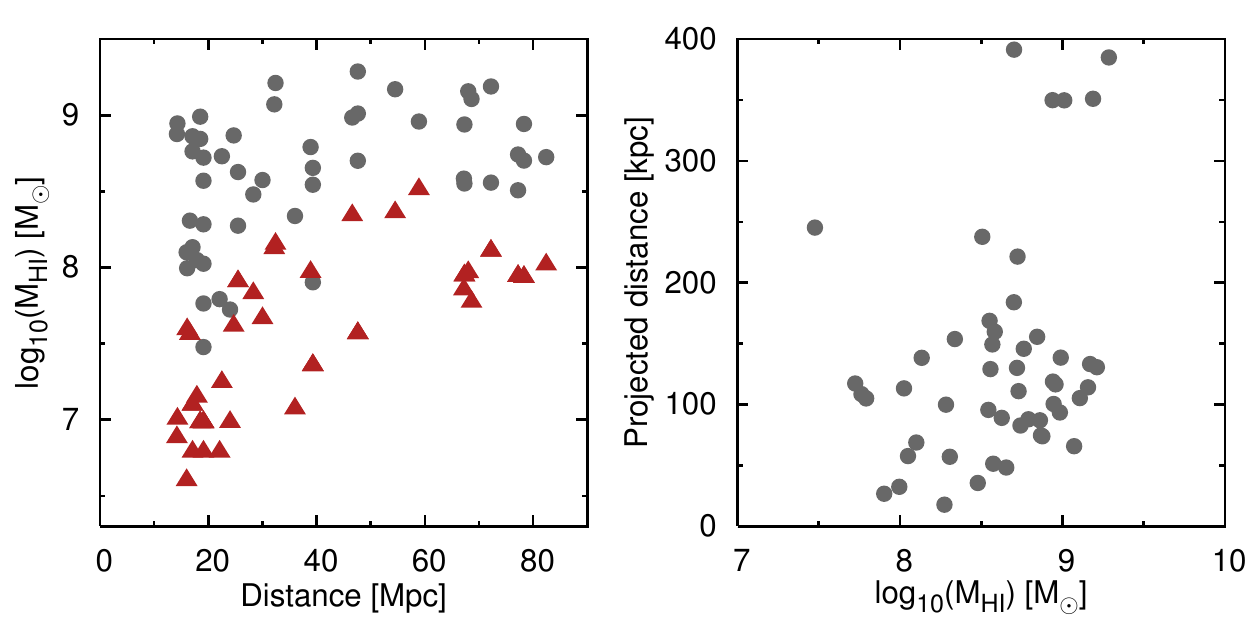}
\caption{ \emph{Left panel}: the gray dots show the \hi\ masses of detected companions as a function of distance, red triangles are the lowest detectable mass in the correspondent data-cubes. \emph{Right panel}: distance from the main galaxies projected onto the plane of the sky of the detected satellites as a function of their \hi\ mass.}
\label{fig:masses}
\end{figure*}

We found that, among 148 data-cubes, 101 ($\sim 68.2\%$) had no detectable companions, whereas 47 ($\sim 31.8 \%$) contained multiple systems. Among these 47 data-cubes, 15 ($\sim 10.1 \%$ of the total, $\sim 31.9 \%$ of multiple systems) contained only galaxies with similar masses ($M_\mathrm{bar, sat}/M_\mathrm{bar, main}>0.20$). Six data-cubes ($\sim 4.1 \%$ of the total, $\sim 12.8 \%$ of multiple systems) show both major and minor companions and 26 data-cubes ($\sim 17.6 \%$ of the total, $\sim 55.3 \%$ of multiple systems) show only dwarf companions. Overall, among 148 analysed data-cubes, 21 ($\sim 14.2 \%$), show companions which could be possible candidates for a future major merging, while 32 data-cubes ($\sim 21.6 \%$) show potential candidates for minor mergers. Some examples of spiral galaxies with minor satellites are shown in Fig. \ref{fig:examples}.

We focused on potential minor mergers and all data-cubes with only major companions were excluded from the further analysis. For the six data-cubes with both type of companions, we assumed as the main galaxy the one with the largest \hi\ mass and we ignored the other spiral galaxies. The total number of dwarf gas-rich satellites detected is 50 (Tab. \ref{tab:galaxies}). Forty-six dwarf galaxies have a clear optical counterpart in the Sloan Digital Sky Survey (SDSS) or in the Digitized Sky Survey (DSS) images. Four detections, marked with an asterisk in  Tab. \ref{tab:galaxies}, are not univocally identifiable and they could be either very faint dwarf galaxies or \hi\ clouds. Most satellites are already catalogued in galaxy archives. Ten galaxies, marked with a dagger in Tab. \ref{tab:galaxies},  seem not to be catalogued. The \hi\ masses of the detected dwarf galaxies vary between about $10^7\,$ M$_\odot$ and few $10^9\,$ M$_\odot$. The \hi\ mass distribution of the main galaxies and their minor satellites is shown in Fig. \ref{fig:MF}. The mass function for spiral galaxies is peaked at $\log \mathrm{M_{HI \, [M_\odot]}}\sim9.5$, consistently with studies on wider \hi\ samples \citep[e.g.,][]{Zwaan+05}. Most dwarf companions have masses of a few $10^8 \; \mathrm{M}_\odot$ and their mass distribution has a cut-off above $5 \times 10^9 \; \mathrm{M}_\odot$. This is partially due to our selection criteria. However, it is interesting to note that this distribution is fairly comparable with that of \hi -rich dwarf galaxies in the Local Group and in Local Group analogues \citep[e.g.,][]{Grcevich&Putman09, Pisano+11}. In Fig. \ref{fig:masses} we show the \hi\ masses of the detected dwarf galaxies as a function of the distance from the Milky Way. The red triangles represent the minimum detectable mass for each data-cube, calculated using equation (\ref{eq:HImass}) on a three-dimensional region with the size of a spatial beam times the velocity resolution (two channels) and a flux of $4\times$R.M.S. noise of the cube. This is the minimum mass that an object must have to be accepted by the source-finding algorithm. Note the bias effect on the detectable mass due to the distance (see discussion in \ref{sec:uncertainties}).

The projected distances of the dwarf satellites from the main galaxies usually range from some dozen to a few hundred kiloparsecs and typical time-scales for collisions, estimated through the parabolic orbit approximation, are between  $\sim100$ Myr and 2 Gyr. The number of dwarfs within 100 \kpc\ from the main galaxies and between 100 and 200 \kpc\ is almost the same. In the \emph{right panel} of Fig. \ref{fig:masses} we show the projected distance as a function of the dwarf \hi\ masses. Within 200 \kpc , dwarf galaxies are quite uniformly distributed over the \hi\ masses. There is a weak tendency for companions to be more massive at larger distances, as we may expect. However, there is an observational bias that can affect this plot. It is a combination of two effects: the linear  field-of-view of the  observations increases with distance, while the minimum detectable mass (Fig. \ref{fig:masses}, \emph{left panel}) and the linear resolution decrease. Thus we may detect preferentially companions with lower masses closer to the main galaxies and vice-versa. Moreover, there is also a selection effect due to the primary beam attenuation, i.e. at large angular distances, only massive systems are detected because of the lower sensitivity of the instrument. These effects make it difficult to compare our findings with studies of dwarfs galaxies in the Local Group.

\begin{table*}[!t]
\small
\caption{\normalsize Detected companions of the WHISP spiral galaxies with $M_\mathrm{bar,sat}/M_\mathrm{bar,main} \le 0.20$: (1) First name in NED archive or DF if not classified, (2) UGC name of the main galaxy, (3) celestial coordinates, (4) adopted distance [same as the main galaxy or taken from EDD catalogue, when specified], (5) systemic velocity,  (6) line width of the global profile at the 20\% level, (6) total \hi\ mass, (8) projected distance from the main galaxy (9) time of collision with the main galaxy in a parabolic orbit, (10) gas accretion rate onto the main galaxy.}
\label{tab:galaxies} 
\centering
\begin{tabular}{llcccccccccc}
\hline\hline\noalign{\vspace{5pt}}
Name 	& Main galaxy 	& Coord. (J2000) 	& $D$ & $v_\mathrm{sys}$ & $w_{20}$ & $M_\mathrm{HI}$ & $d_\mathrm{proj}$ & $t_\mathrm{coll}$ & $\dot{M}_\mathrm{HI}$ \\

&	& RA-Dec & Mpc & km/s & km/s & $10^8 \, \mo$ & kpc & $10^8$ yr & $\, \mo$/yr \\
\hspace{5pt} (1) & \hspace{15pt} (2) & (3) & (4) & (5) & (6) & (7) & (8) & (9) & (10)\\
\noalign{\smallskip}
\hline\noalign{\vspace{5pt}}
AGC 102802 	&	 UGC 485   	&	J004702.7+301243	&	58.9	&	5296	&	85	&	9.08	&	117	&	15.7	&	0.58	\\
AGC 113996 	&	 UGC 624   	&	J010107.2+304052	&	78.3	&	4762	&	29	&	8.76	&	119	&	11.6	&	0.75	\\
AGC 113884 	&	 UGC 624   	&	J010000.3+302357	&	78.3	&	4717	&	96	&	5.04	&	391	&	17.8	&	0.28	\\
$[$VH2008$]$ J0101+4744 	&	 UGC 625   	&	J010118.4+474432	&	28.3	&	2795	&	62	&	3.02	&	36	&	4.5	&	0.67	\\
DF1$^\dagger$	&	 UGC 1437  	&	J015708.1+354825	&	54.5	&	4592	&	172	&	14.80	&	133	&	9.8	&	1.51	\\
PGC 9994      	&	 UGC 2141  	&	J030653.0+301542	&	24.7	&	812	&	43	&	7.37	&	75	&	5.6	&	1.32	\\
PGC 2328690 	&	 UGC 2459	&	J030225.7+485452	&	32.4	&	2449	&	134	&	16.29	&	131	&	14.3	&	1.14	\\
$[$KLT2208$]$ HI J0302+352$^*$ &	 UGC 2487  	&	J030210.5+351627	&	72.2	&	4933	&	55	&	15.43	&	351	&	18.1	&	0.85	\\
$[$SOS2010$]$ J0301491+3529012 	&	 UGC 2487  	&	J030147.2+352839	&	72.2	&	4876	&	38	&	3.61	&	129	&	8.6	&	0.42	\\
UGC2813 	&	 UGC 2800  	&	J034234.1+711828	&	16.1$^1$	&	1381	&	62	&	0.99	&	33	&	4.2	&	0.24	\\
HFLLZOA G136.96+14.21 	&	 UGC 2916  	&	J040403.5+713707	&	68.0	&	4450	&	158	&	14.36	&	114	&	10.8	&	1.33	\\
2MASX J04550438+3002212 	&	 UGC 3205  	&	J045826.3+295653	&	47.6	&	3239	&	173	&	10.26	&	350	&	17.6	&	0.58	\\
DF2$^\dagger$	&	 UGC 3205  	&	J045504.2+300209	&	47.6	&	3530	&	47	&	5.03	&	184	&	9.7	&	0.52	\\
DF3$^\dagger$	&	 UGC 3205  	&	J045653.8+293602	&	47.6	&	3229	&	110	&	19.34	&	385	&	18.3	&	1.06	\\
DF4$^\dagger$	&	 UGC 3382  	&	J055903.3+621719	&	67.2	&	4407	&	64	&	3.83	&	160	&	12.6	&	0.30	\\
DF5$^\dagger$	&	 UGC 3407  	&	J060841.0+415647	&	39.3	&	3683	&	66	&	3.50	&	96	&	10.1	&	0.35	\\
DF6$^\dagger$	&	 UGC 3407  	&	J060913.3+420104	&	39.3	&	3688	&	114	&	4.51	&	48	&	7.9	&	0.57	\\
DF7$^\dagger$	&	 UGC 3407  	&	J060853.9+420338	&	39.3	&	3693	&	73	&	0.80	&	27	&	3.5	&	0.23	\\
DF8$^{\dagger\,*}$	&	 UGC 3422  	&	J061633.1+705743	&	77.2	&	4009	&	24	&	3.22	&	238	&	9.8	&	0.33	\\
GALEXASC J061256.68+710650.6 	&	 UGC 3422  	&	J061254.8+710659	&	77.2	&	3998	&	104	&	5.52	&	83	&	7.5	&	0.74	\\
NPM1G +60.0018 	&	 UGC 3546  	&	J065150.2+604122	&	17.9	&	1768	&	52	&	1.12	&	58	&	5.8	&	0.19	\\
GALEXASC J070643.91+635521.0	&	 UGC 3642  	&	J070645.1+635515	&	67.3	&	4714	&	106	&	3.56	&	169	&	11.1	&	0.32	\\
UGC 3660 	&	 UGC 3642  	&	J070634.1+635056	&	67.3	&	4261	&	75	&	8.70	&	350	&	17.9	&	0.49	\\
KUG 0829+227B 	&	 UGC 4458  	&	J083247.7+223443	&	68.6	&	4621	&	231	&	12.80	&	105	&	12.4	&	1.03	\\
MCG +10-13-030 	&	 UGC 4666  	&	J085422.1+585908	&	16.0	&	1016	&	90	&	1.26	&	69	&	5.7	&	0.22	\\
SDSS J091001.72+325659.8 	&	 UGC 4806  	&	J091005.0+325607	&	25.5	&	2049	&	125	&	4.23	&	89	&	9.5	&	0.45	\\
KUG 0906+333A 	&	 UGC 4806  	&	J090919.5+330734	&	25.5	&	1897	&	60	&	1.88	&	18	&	5.7	&	0.33	\\
SDSS J093137.13+292533.3 	&	 UGC 5060  	&	J093138.0+292534	&	24.0	&	1608	&	77	&	0.53	&	117	&	9.9	&	0.05	\\
KDG 059 	&	 UGC 5253  	&	J095156.6+720439	&	16.6	&	1121	&	46	&	2.03	&	57	&	6.2	&	0.33	\\
UGC 6797 	&	 UGC 6778  	&	J114940.5+482533	&	17.1	&	962	&	81	&	7.28	&	87	&	8.5	&	0.86	\\
SDSS J115027.42+490105.9	&	 UGC 6778	&	J115027.4+490106	&	17.1	&	1120	&	31	&	1.67	&	138	&	11.2	&	0.12	\\
UGC 6791	&	 UGC 6786	&	J114923.6+264428	&	22.5$^1$	&	1866	&	274	&	5.38	&	111	&	10.0	&	0.54	\\
SDSS J114820.16+562045.7 	&	 UGC 6787  	&	J114820.6+562049	&	22.1	&	1080	&	28	&	0.62	&	105	&	7.6	&	0.08	\\
UGC 6733 	&	 UGC 6787  	&	J114535.7+555313	&	19.1$^2$	&	1158	&	187	&	5.26	&	130	&	10.3	&	0.51	\\
UGC 6816 	&	 UGC 6787  	&	J115047.5+562719	&	17.1$^1$	&	887	&	115	&	5.78	&	146	&	11.0	&	0.52	\\
SDSS J122442.59+544441.3 	&	 UGC 7506  	&	J122440.2+544448	&	36.0	&	2495	&	109	&	2.18	&	154	&	11.6	&	0.19	\\
UGC 8005	&	 UGC 7989  	&	J125149.1+254644	&	14.3$^1$	&	1196	&	198	&	8.84	&	101	&	8.6	&	1.02	\\
UGC 8254 	&	 UGC 8307  	&	J131038.2+363807	&	19.1	&	1088	&	105	&	3.71	&	149	&	16.1	&	0.23	\\
DF9$^{\dagger\,*}$	&	 UGC 8307  	&	J131153.6+362758	&	19.1$^1$	&	954	&	75	&	1.92	&	100	&	14.2	&	0.14	\\
UGC 8271 	&	 UGC 8307  	&	J131131.3+361655	&	18.5$^1$	&	1145	&	150	&	6.99	&	156	&	22.1	&	0.32	\\
DF10$^{\dagger\,*}$ 	&	 UGC 8307  	&	J131134.3+362942	&	19.1	&	1191	&	32	&	0.58	&	109	&	15.5	&	0.04	\\
KUG 1309+362 	&	 UGC 8307  	&	J131146.7+355731	&	19.1	&	1123	&	26	&	0.30	&	245	&	24.9	&	0.01	\\
UGC 8303 	&	 UGC 8307  	&	J131317.6+361303	&	18.5$^1$	&	948	&	92	&	9.77	&	139	&	20.3	&	0.48	\\
UGC 8314 	&	 UGC 8307  	&	J131401.0+361908	&	19.1	&	938	&	71	&	1.06	&	113	&	21.2	&	0.05	\\
MCG +08-27-001 	&	 UGC 9366  	&	J143359.2+492647	&	38.9	&	2122	&	127	&	6.18	&	88	&	6.2	&	1.00	\\
KUG 1512+557	&	 UGC 9797  	&	J151400.2+553222	&	46.6	&	3550	&	154	&	9.66	&	94	&	9.4	&	1.03	\\
SDSS J152617.51+404004.0 	&	 UGC 9858  	&	J152617.9+404008	&	32.2	&	2687	&	51	&	11.80	&	66	&	7.7	&	1.53	\\
MCG +08-34-005 	&	 UGC 11283	&	J183400.4+492233	&	30.0	&	2076	&	63	&	3.75	&	51	&	7.5	&	0.50	\\
GALEXASC J215645.61+275419.5	&	 UGC 11852	&	J215645.7+275418	&	82.4	&	5710	&	46	&	5.30	&	221	&	15.2	&	0.35	\\
ZOAG G095.92-08.72	&	 UGC 11951	&	J221145.4+453649	&	14.2	&	1145	&	78	&	7.52	&	74	&	8.3	&	0.91	\\

\noalign{\vspace{2pt}}\hline
\noalign{\vspace{5pt}}
\multicolumn{3}{l}{$^\dagger$ Not catalogued in the NED, HyperLEDA or SIMBAD archives. } \\
\multicolumn{8}{l}{$^*$ Without a clear optical/UV counterpart in DSS, SDSS or GALEX images. DF8 is not covered by the SDSS survey.} \\
\multicolumn{3}{l}{$^1$ Distance from EDD catalogue.} \\
\end{tabular}
\end{table*}

\begin{figure*}[t]
\centering
\includegraphics[width=0.80\textwidth]{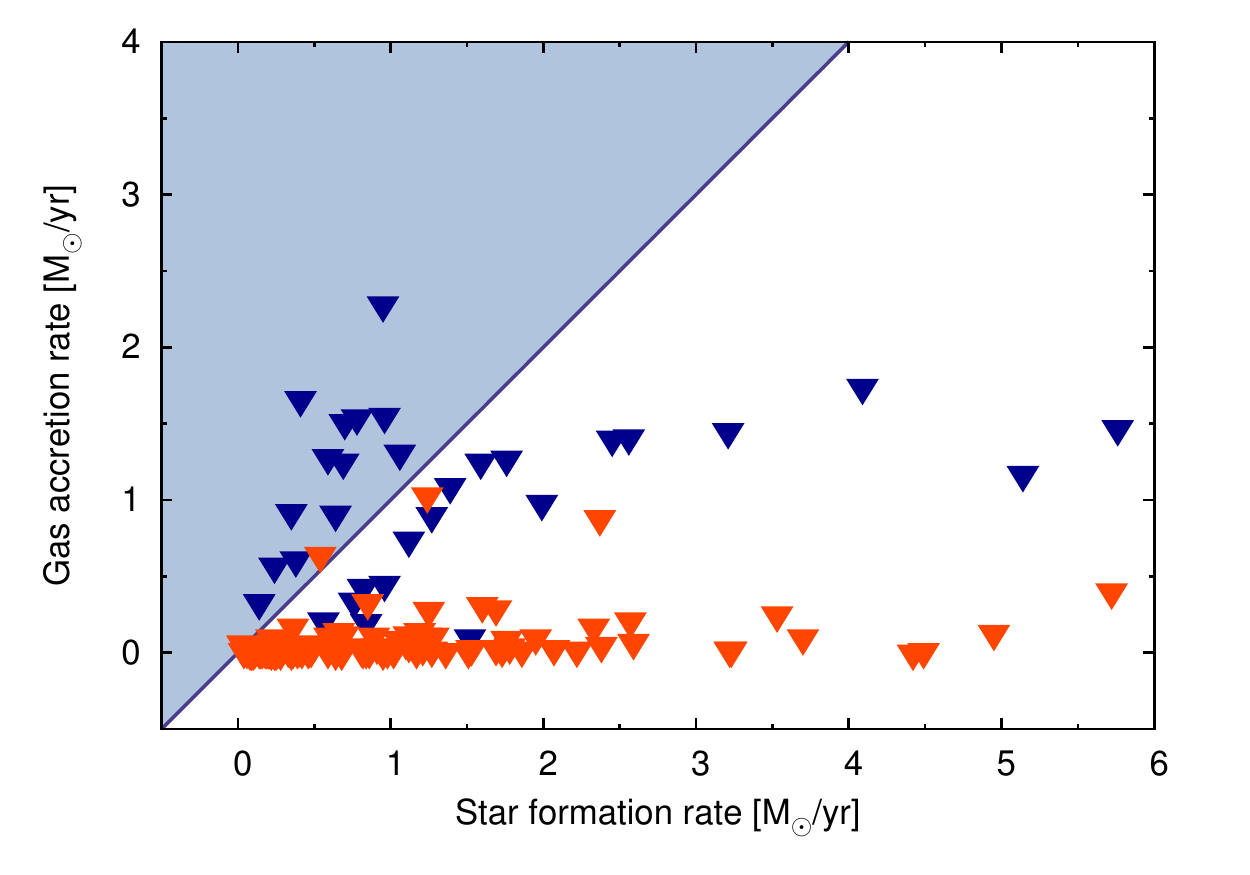}
\caption{ Upper limits to the cold gas accretion rates from satellites vs star formation rates in spiral galaxies in the WHISP sample. The inverted dark blue triangles are upper limits to the gas accretion rate for galaxies with detected satellites (including both visible and hidden accretion), the inverted orange triangles are the hidden accretion upper limits for galaxies without companions, estimated as discussed in the text. The star formation rates are lower limits calculated from the far-infrared fluxes. The blue-shadowed region represents a complete feeding of SF through minor mergers.}
\label{fig:sfr-acc}
\end{figure*}

The systemic velocity of dwarf galaxies is calculated as the average midpoint between the velocities at the 20\% and 50\% of the peak flux of their global \hi\ profiles. The $\Delta v_\mathrm{sys} = \,\,\parallel v_\mathrm{sys, main}-v_\mathrm{sys, sat}\parallel\,$ ranges between a few tens to a few hundreds $\kms$. Satellites do not have systemic velocities that differ more than 300 $\kms$ from those of the main galaxies. The velocity widths $w_{20}$ of dwarf galaxies, taken at the 20\% of the peak flux, are usually lower than 200 $\kms$, even if corrected for a mean inclination of 60 degrees, except for three galaxies. Overall, most of the satellites have $w_{20}<100$ $\kms$.

For each data-cube with identified dwarf companions, we calculated the maximum possible accretion rate of cold hydrogen gas $\dot{M}_\mathrm{HI}$ onto the main galaxy, the star formation rate $\dot{M}_\mathrm{SF}$ of the main galaxy and the ratio $\dot{M}_\mathrm{HI}/\dot{M}_\mathrm{SF}$. For all galaxies, with or without identified companions, a potentially hidden accretion from dwarfs below the detectability limit was estimated. The hidden accretion rate was calculated by dividing the above-mentioned minimum detectable mass by the average collision time over the sample, i.e $1.1 \, \Gyr$. Integrating the \hi\ mass function ($\phi_*[\mathrm{Mpc^{-3}\,dex^{-1}}\,h_{70}^3] = 4.8 \pm 0.3 \times 10^3 $, $\log(M_*/M_\odot)+2\log h_{70}=9.96\pm0.02$ and $\alpha = -1.33\pm0.02$, \cite[][]{Martin+10}) below the detection limit and within the volume of each data-cube always gives \hi\ masses lower than minimum detectable mass. Thus, with our choice we are maximizing the mass of the undetected galaxies.

We found a mean upper limit for the accretion in galaxies with identified minor companions of 0.86 $\moyr$, with a mean ratio $\langle \dot{M}_\mathrm{HI}/\dot{M}_\mathrm{SF} \rangle \sim 0.67$. A more meaningful estimate is however the mean upper limit to the accretion over the whole sample, that turns out to be $\dot{M}_\mathrm{HI}=0.28 \, \moyr$ against the average star formation rate of 1.29 $\moyr$, with a mean ratio $\langle \dot{M}_\mathrm{HI}/\dot{M}_\mathrm{SF} \rangle \sim 0.22$. The median of $\dot{M}_\mathrm{HI}/\dot{M}_\mathrm{SF}$ is 0.07. Thus, the ratio of the gas needed for star formation to the maximum gas accretion provided by minor mergers is between 5 and 14. Considering a fraction of gas recycle from stellar feedback of 30\% \citep[e.g.][]{Naab&Ostriker06} leads to a ratio between 3 and 10.

The above results show that the number of dwarf galaxies in the Local Universe is on the average too low to guarantee the continuous gas replenishment needed by star formation. In Fig. \ref{fig:sfr-acc} we show a plot of $\dot{M}_\mathrm{HI}$ versus $\dot{M}_\mathrm{SF}$ for each galaxy individually. If the gas accretion were large enough to sustain the star formation of the main galaxies, the data points would have fully populated the blue-shadowed region in the upper-left corner, whereas the vast majority lie well below the blue straight-line indicating a ratio $\dot{M}_\mathrm{HI}/\dot{M}_\mathrm{SF}=1$. 
We conclude that minor mergers can not bring enough gas to the discs and sustain star formation. Once again, our values of gas accretion rates are very strong upper limits, because of our very stringent assumptions, and the real accretion rates could reasonably be one order of magnitude lower than our estimate. Incidentally, we note that our assumptions would imply that all dwarf galaxies disappear in the next 2 Gyr. We stress that our SFRs are likely lower limits as they are calculated using only FIR fluxes. This bias goes in the direction of strengthening our findings. 

We repeated the analysis of the WHISP data-cubes using a sigma-clipping threshold for the sources of $3\sigma$ and $5\sigma$. Reducing the detection threshold leads the program to identify many more dwarf companions: more than 100 minor satellites are detected at the lower level, but most of these sources are clearly false detections and the results obtained would be very likely unreliable. Such a large fraction of wrong detections is probably due to the low signal-to-noise of the WHISP data-cubes. Instead, increasing the detection threshold to 5$\sigma$ leads to results very similar to those described above as just two of the dwarf companions found at $4\sigma$ are missed by the rejection criteria, namely the satellites of UGC 7506 and UGC 9858. These companions are actually good detections, as quoted in the literature \citep{Noordermeer+05}, but at a $5\sigma$ level they are discarded by the one beam covering requirement. The mean values of the accretion rate at $5\sigma$ are also in agreement with those found at $4\sigma$.

\section{Discussion}\label{sec:discussion}

The application of our code to the WHISP catalogue led to a firm upper limit for the accretion of cold gas from minor mergers in the Local Universe of $ 0.28 \;\moyr$. The total multiple system fraction for the WHISP sample is $\sim 32\%$, in particular $\sim 22\%$ of galaxies are accompanied by minor companions and $\sim 14\%$ are major systems. Here we discuss the main uncertainties of our results and their relevance.

\subsection{Uncertainties}\label{sec:uncertainties}

Our estimate does not take into account the molecular fraction. The amount of molecular gas in dwarf galaxies is highly unconstrained as they are often undetected in CO emission lines \citep[e.g.,][]{Taylor+98}. They also usually have low metallicities, making the conversion between CO and H$_2$ even more uncertain \citep[e.g.,][]{Boselli+02}. However, any realistic correction for molecular gas should not increase our accretion rate by more than a factor two. 

WHISP is a source-targeted survey and it can not be obviously considered as a complete sample. The selection criterion, grounded on the apparent size of the observed galaxies, produces a catalogue that favours progressively larger and more massive galaxies moving to greater distances from the Milky Way. This effect can be appreciated in Fig. \ref{fig:masses} (upper envelop) although it appears to be not too severe.  The growth with the distance of the minimum detectable mass furthermore makes it impossible to detect low mass satellites at large distances. In order to test the importance of these biases, we have considered only those data-cubes with a minimum detectable mass $M_\mathrm{det} \leq 10^8 \mo$. In this way, we can obtain a sub-sample of galaxies where satellites are quite uniformly distributed over the mass and the distance ranges (\emph{left panel} of Fig. \ref{fig:masses}). The maximum accretion rate obtained in this case is  $0.21 \moyr$. Reducing the threshold to data-cubes with $\log M_\mathrm{det} \leq 5\times10^7 \mo$ leads to a maximum accretion rate of $0.18 \moyr$. These values indicate that our accretion rate estimates is not strongly affected by the incompleteness of the sample of the dwarf galaxies.

Another bias effect is related to the linear field of view, which is greater at larger distances. In the farthest systems,  the field of view allows us to observe satellites with projected distances of some hundreds kpc from the main galaxies, whereas we can not go beyond one hundred kpc in the closest systems. The primary beam attenuation of the WSRT is significantly large ( $\sim80\%$ of the flux is missed) beyond $25'$ from the pointing center, corresponding to $\sim70$ kpc at about 10 Mpc.  This indicates that we should be able to detect fairly separated satellites also in the nearest systems. The most distant satellites have larger collision time-scales and their contribution to the global accretion is expected to be smaller. In our sample, considering only satellites within 100 kpc from the main galaxies gives an accretion rate of 0.38 $\moyr$,  0.27 $\moyr$ between 100 and 200 kpc and 0.21 $\moyr$ beyond 200 kpc (the global value being 0.86 $\moyr$) . These results show that the contribution of very distant satellites is progressively less important, thus the limited field of view of the closest systems should not significantly affect our accretion rate estimate.

 In the literature, mergers are usually classified on the basis of their dynamical mass ratio: pair of galaxies with $M_\mathrm{sat}/M_\mathrm{main} \le 0.1-0.2$ are considered minor mergers, otherwise major mergers. Unfortunately, we can not trivially estimate the dynamical masses of satellite galaxies from the \hi\ data. Thus, in this work, we divided satellites depending on the ratio of their baryonic mass to the main galaxy baryonic mass. Satellites with baryonic content lower than 20\% of the main galaxies ($M_\mathrm{bar,sat}/M_\mathrm{bar,main} \le 0.20$) are classified as minor companions. This is an arbitrary but conservative choice, since most detected satellites have mass ratio $\ll 0.05 $. It is however interesting to quantify the accretion rate using different baryonic mass ratios. In our sample, the maximum accretion rate ranges between $0.20 \moyr$ for $M_\mathrm{bar, s}/M_\mathrm{bar,g} \le 0.1$ and $0.56 \moyr$ for $M_\mathrm{bar,sat}/M_\mathrm{bar,main} \le 0.5$. If we consider the whole galaxy pairs as potential mergers and we calculate the accretion rate by accreting the less massive ones onto the most massive ones, we obtain the value of $1.22 \moyr$. Even such an excessive overestimate turns out to be of the same order of the mean SFR. We conclude that mergers in the Local Universe can not sustain the star formation in spiral galaxies.

\subsection{Comparison to other estimates}

The accretion of cold gas from minor mergers in the Local Universe has been estimated by \cite{Sancisi+08}, visually inspecting and comparing total maps, velocity fields and position-velocity diagrams for the WHISP galaxies. They found a minor merger fraction of about 25\%. Unlike our approach, they considered only those systems that show clear signs of tidal interactions, such as tails, bridges, disturbed \hi\ morphologies and/or kinematics. Assuming typical \hi\ masses of the  dwarfs of the order $10^{8-9} \; \mathrm{M}_\odot$ and a lifetime for observed features of about 1 Gyr, Sancisi et al. inferred a mean accretion rate of \hi\ gas around $0.1-0.2 \; \moyr$ and they stressed that such a value is likely a lower limit. It is worth noting that most systems we considered as potential minor mergers were not recognized that way by Sancisi et al. and, on the contrary, many interactions they identified were not found by our code. The reason is simple: our code looks for ``separated'' objects and it handles all dwarf companions as candidates for minor mergers, also those showing no signs of ongoing interaction. In other words, we look at the population of dwarfs in the environment of a spiral galaxy that could become a minor merger in the next future. Our code identifies companions until the two galaxies start ``touching'' each other and we estimate the accretion rate using the time-scale for collision as accretion time. Instead, \cite{Sancisi+08} find a later stage of merging, i.e., when galaxies are strongly interacting and the gas is visibly disturbed in the morphology and/or in the kinematics. Consequently, they calculate the accretion rate using as time-scale the dynamical time that it should take for these features to disappear as the gas redistributes uniformly in the disc. In our work the accretion process ends when galaxies touch each other, whereas for \cite{Sancisi+08} that is the starting point. However, since the population of dwarf galaxies has likely remained similar in the last Gyr or so, the two accretion rates should be comparable. Interestingly, our upper limit of $ \dot{M}_\mathrm{HI}< 0.28 \;\mathrm{M}_\odot \; \mathrm{yr}^{-1}$ is not in contradiction with the average accretion rate estimated by \cite{Sancisi+08}. 

\subsection{Merger fraction}
Most of recently published studies on the local merging systems have been made using images from optical-UV galaxy surveys \citep[e.g.,][]{Patton+00, Lambas+12, Robotham+12} such as the Second Redshift Survey of Southern Sky (SRSS2), the Sloan Digital Sky Survey (SDSS) and the recent Galaxy And Mass Assembly (GAMA) survey, whereas just a couple of studies have been carried on using \hi\ data \citep{Sancisi+08, Holwerda+11}. These studies have mainly investigated the fraction and the rate  (fraction of mergers per comoving volume and time units) of galaxies showing signs of interactions and their evolution with time. 

To date, two main approaches have been used to estimate the galaxy merger fraction and both make use of high resolution imaging. The pair method consists in counting the galaxies spatially separated from each other by less than a few tens of kpc and with spectroscopic radial velocities that do not differ by more than a few hundreds of $\kms$ \citep[e.g.,][]{LeFevre+00, Lin+08}. Using this kind of approach it is possible to estimate a ``progenitor galaxy'' merger fraction. 
The second approach identifies mergers by quantifying morphological signatures that can be related to past or ongoing interactions, such as asymmetries and/or tails. This method makes use of several parameters for describing peculiar light distributions, such as the Concentration-Asymmetry-Smoothness parameters \citep[CAS,][]{Conselice03} or the Gini-M$_{20}$ parameters \citep{Lotz+04}. This technique can identify mergers in a relatively late stage, but not all asymmetric galaxies are necessary merger features. 
The asymmetry method is similar to the technique used by \cite{Sancisi+08}, whereas our approach on \hi\ data-cubes is conceptually similar to the close pair method.  The main difference is that we do not impose any limit for the projected distance between galaxies, whereas the velocity criterion is implicit in the data-cubes. Moreover, we select objects in 3D space (so potentially also overlapping in the sky) and we estimate the minimum time of collision for each galaxy independently.

The asymmetry and close pairs methods have been widely used with optical galaxy surveys, but, despite the large number of studies, there is little consensus on the galaxy merger rate and its evolution with redshift. Current observations of the fraction of galaxy undergoing a merger differ by an order of magnitude, from $\sim2\%$ \citep[e.g.,][2.3\% and 1.9\%, respectively]{Patton+00, DePropris+07} to 15\% \citep[e.g.,][]{deRavel+09} and its trend with redshift vary from no evolution \citep[e.g.,][]{Jogee+09} to strong evolution \citep[e.g.,][]{Lopez-Sanjuan+09}. These discrepancies mainly arise from the different criteria for galaxy counting, merger selection and bias in the galaxy samples. The value we found ($\sim 32\%$) is a companion fraction rather than a merger fraction as some companions that we considered are fairly far away from the main galaxies (Fig. \ref{fig:masses}, \emph{right panel}). It is therefore difficult to compare our fraction with the above mentioned values. Broadly speaking, our estimate, which is indeed an upper limit, is higher at least of a factor 2-3 because our program treats all multiple systems as mergers and, working with \hi\ data, identifies more easily dwarf gas-rich companions compared to optical observations. However, if we exclude the very far away companions, namely those beyond 100 \kpc\ of projected distance, we obtain a companion fraction of $\sim 14\%\,$, not too different from the values found with optical studies. Finally, we stress that the WHISP sample is insignificant compared to other local references based on large catalogues, such as the SDSS or the Millennium Galaxy Catalogue (MGC), so that our values are less reliable from a statistical point of view.
 
A recent study carried out by \cite{Holwerda+11} estimated the merger fraction and rate for the whole WHISP sample using both close pair and asymmetry methods on \hi\ total maps. Holwerda et al. found a merger fraction of 7\% based on pairs, and 13\% based on disturbed morphology. We can not compare our merger fraction with the latter value, because our program ignores the galaxy morphology, but the former value is fully comparable and our estimate is significantly higher by about a factor 4. A possible reason of such a discrepancy is that Holwerda et al. based their pair fraction on 24 multiple systems previously identified and classified as interacting by Noordermeer et al. (2005b) and Swaters et al. (2002b), whereas our code detected a much larger number of satellites (see Tab. \ref{tab:galaxies}). If we use this sub-sample, the merger fractions become closely comparable. 

\subsection{Other channels for gas accretion}

How star-forming galaxies can sustain their star formation is still an open question. In this study, we demonstrated that gas-rich minor mergers do not play a primary role and other dominant accretion channels must be admitted. A way to fill the discrepancy between the estimated accretion rates and the SFRs could be to assume that the \hi\ mass function were much steeper in the recent past than now, so that the number of dwarf satellites to be accreted were much higher. However, to date no observational evidence in that direction can be achieved with the present generation of radio-telescopes and studies of the Damped Lyman $\alpha$ systems show a remarkable constancy of the \hi\ mass throughout the Hubble time \citep[e.g.,][]{Prochaska&Wolfe09}. Another possibility is that the most accretion is supported by infalling of gas clouds with \hi\ masses of $10^7-10^6 \, \mo$, but recent deep observations of nearby groups of galaxies \citep[e.g.,][]{Pisano+07,  Chynoweth+09}, as well as large blind surveys, such as ALFALFA \citep{Giovanelli+07}, showed no evidence for a significant population of these small \hi\ clouds. Moreover, studies on the Milky Way's High Velocity Clouds (HVCs) estimated a contribution to the total gas accretion of $0.1-0.2 \moyr$ \citep[e.g.,][]{Wakker+07, Putman+12}, a value much smaller than the SFR. 
In addition, the gas in the ionized phase could produce a further accretion rate of $\sim 1 \moyr$ \citep[e.g.,][]{Shull+09}, but it is not understood whether this gas can feed the star formation process in the disc. Numerical simulations \citep[e.g.,][]{Fernandez+12} support the idea that the most of the gas infall in Milky Way-like galaxies is continuously provided by a drizzle and filamentary cosmological accretion, which would be almost undetectable or very difficult to identify \citep[e.g.,][]{Lehner+13, Tumlinson+13}. Finally, large amounts of matter could be supplied by the coronal gas cooling potentially triggered by supernova feedback \citep{Marinacci+10}.

\section{Conclusions and future prospects}

In this paper, we estimated the maximum accretion of cold gas from minor mergers in a sample of large spiral galaxies from the WHISP catalogue. We used an algorithm of source finding to detect dwarf \hi -rich satellites around these spiral galaxies and we assumed that they will disappear and merge with the main galaxies in the shortest possible time. We found that $\sim 22\%$ of galaxies have detected dwarf companions ($M_\mathrm{bar,sat}/M_\mathrm{bar,main} \le 0.20$) and we estimated a maximum gas accretion rate onto the main galaxies over the whole sample of  $0.28 \moyr$. Given the assumptions, this value is a strong overestimate and the actual value can easily be an order of magnitude or more lower. From far-infrared luminosities, we calculated a mean star formation rate of $1.29 \moyr$, a value which is nearly five times higher than the maximum gas accretion rate. These results strongly suggest that minor mergers can not bring enough gas to guarantee a long lasting star formation process in the discs of the spiral galaxies. We note that our method can also detect, if present, large floating \hi\ clouds and include them in the accretion budget. We did not find any significant population of these clouds. Thus, most of gas accretion seems to be hidden to the current investigations in \hi\ emission.

WHISP is a fairly large sample of nearby galaxies, but it is very small compared to surveys carried out at other wavelengths. The new generation of radio telescopes, such as the SKA \citep{Carilli&Rawlings04} and its pathfinders, ASKAP \citep{Johnston+08} and MeerKAT \citep{Booth+09} and the restilying of existing interferometers, such as the WSRT with the APERTIF system \citep{Verheijen+08} and the Karl G. Jansky VLA, will largely increase the number of available data samples. In the next future, already scheduled \hi\ surveys, such as WALLABY and DINGO with ASKAP, LADUMA with MeerKAT and WNSHS with WSRT/APERTIF, will increase the number of galaxies observed with radio interferometers by three orders of magnitude, from a few hundreds to about $10^5$. It will be very interesting to apply the kind of analysis performed in this paper to those large galaxy samples.

\begin{acknowledgements}
We thank Renzo Sancisi, Tom Oosterloo, Thijs van der Hulst and Micol Bolzanella for helpful suggestions and fruitful discussions. EdT personally thanks Gabriele Pezzulli for his daily help and useful advices. We used the WHISP data sample and the EDD, NED, HyperLEDA and Simbad catalogues. This research made use of some parts of the Duchamp code, produced at the Australia Telescope National Facility, CSIRO, by Matthew Whiting. We acknowledge financial support from PRIN MIUR 2010–2011, project ‘The Chemical and Dynamical Evolution of the Milky Way and Local Group Galaxies’, prot. 2010LY5N2T.
\end{acknowledgements}

\appendix

\section{Global properties of the main galaxies}

In this Appendix, we list the main properties of the spiral galaxies selected from the WHISP sample for this work. Columns as follows.

{\it Column} (1) gives the UGC name.

{\it Column} (2) provides an alternative common name, like NGC, DDO or IC classifications.

{\it Column} (3) provides the adopted distance in Mpc. We preferably used the EDD catalogue \citep{Tully+09}, otherwise, we used the following distance sources, in the given order: Cosmicflows-2 \citep{Tully+13}, NED archive, Hubble flow with $H_0=70$ km s$^{-1}$ Mpc$^{-1}$ and systemic velocities corrected for Virgo-infall taken from the HyperLEDA catalogue.

{\it Columns} (4) and (5) give the radius R$_{25}$, namely the length of the projected semi-major axis of a galaxy at the isophotal level 25 mag arcsec$^{-2}$ in the B-band. R$_{25}$ is taken from the HyperLEDA catalogue. In Col. (4) the radius is in arcminutes, in Col. (5) is converted in kiloparsecs using the distances in Col. (3).

{\it Column} (6) provides the inclination angle derived from the axis ratio in B-band as listed in the HyperLEDA catalogue. 

{\it Column} (7) gives the systemic velocity measured in this work as the average midpoint between the velocities at the 20\% and 50\% of the peak flux of the global \hi -line profile.

{\it Column} (8) gives the \hi -line width at the 20\% of the peak flux of the global \hi -line profile, as calculated in this work.

{\it Column} (9) provides the total \hi\ mass estimated in this work.

{\it Column} (10) gives the adopted total baryonic mass $M_\mathrm{bar}$, calculated as described in section \ref{baryonicmass}. 

{\it Column} (11) provides the star formation rate calculated from the 60 $\mu$m and 100 $\mu$m IRAS fluxes.

{\it Column} (12) gives the total gas accretion rate from minor mergers estimated in this work, including detectable and ``hidden'' accretion.

\onecolumn
\begin{longtable}{llcccccccccc}
\label{tab:maingalaxies}\\
\caption{\mbox {Global properties of the main galaxies selected from the WHISP sample.}} \\
\hline\hline\noalign{\vspace{5pt}}
UGC name 	&  Other name & $D$ & $R_{25}$ & $R_{25}$ & $i$ & $v_\mathrm{sys}$ & $w_{20}$ & $M_\mathrm{HI}$ & $M_\mathrm{bar}$ & SFR & $\dot{M}_\mathrm{HI}$  \\

&	& Mpc & ' & kpc & $^\circ$ & km/s & km/s & $10^9 \, \mo$ & $10^9 \, \mo$ & $\mo$/yr & $\mo$/yr\\
\hspace{5pt} (1) & \hspace{15pt} (2) & (3) & (4) & (5) & (6) & (7) & (8) & (9)  & (10) & (11) & (12) \\
\noalign{\smallskip}
\hline\noalign{\vspace{5pt}}

UGC 00094	&	NGC 0026	&	68.6$^1$	&	0.56	&	11	&	47	&	4587	&	320	&	9.63	&	53.34	&	1.17	&	0.04	\\
UGC 00232	&	-	&	65.3$^2$	&	0.52	&	10	&	51	&	4837	&	275	&	7.61	&	38.40	&	1.95	&	0.10	\\
UGC 00485	&	-	&	58.9$^1$	&	1.15	&	20	&	83	&	5246	&	357	&	21.63	&	45.84	&	1.27	&	0.90	\\
UGC 00528	&	NGC 0278	&	12.0	&	1.17	&	4	&	20	&	640	&	138	&	1.32	&	15.68	&	1.02	&	0.01	\\
UGC 00624	&	NGC 0338	&	78.3$^1$	&	0.87	&	20	&	68	&	4770	&	560	&	15.61	&	173.95	&	5.14	&	1.09	\\
UGC 00625	&	IC 0065	&	28.3	&	1.29	&	11	&	73	&	2628	&	360	&	7.68	&	27.05	&	1.12	&	0.74	\\
UGC 00690	&	-	&	74.5$^1$	&	0.85	&	18	&	46	&	5872	&	325	&	9.61	&	56.80	&	0.54	&	0.64	\\
UGC 00731	&	-	&	12.0	&	0.93	&	3	&	24	&	639	&	143	&	0.88	&	38.54$^4$	&	0.21	&	0.01	\\
UGC 00798	&	IC 1654	&	69.4$^2$	&	0.50	&	10	&	40	&	4898	&	222	&	3.96	&	43.96	&	0.60	&	0.11	\\
UGC 01013	&	NGC 0536	&	62.5$^1$	&	1.48	&	27	&	69	&	5187	&	525	&	8.26	&	109.87	&	1.25	&	0.28	\\
UGC 01256	&	NGC 0672	&	8.3	&	3.54	&	9	&	67	&	431	&	240	&	7.56	&	13.99	&	0.18	&	0.01	\\
UGC 01437	&	NGC 0753	&	54.5$^2$	&	0.69	&	11	&	51	&	4905	&	339	&	11.58	&	83.23	&	4.09	&	1.74	\\
UGC 01550	&	NGC 0801	&	52.2$^1$	&	1.38	&	21	&	78	&	5764	&	470	&	15.86	&	75.68	&	2.33	&	0.17	\\
UGC 01633	&	NGC 0818	&	58.1$^1$	&	1.09	&	18	&	70	&	4258	&	501	&	11.35	&	88.10	&	2.57	&	0.21	\\
UGC 01810	&	-	&	109.8$^3$	&	0.87	&	28	&	69	&	7578	&	602	&	31.64	&	210.16	&	2.37	&	0.88	\\
UGC 01856	&	-	&	41.3$^2$	&	1.07	&	13	&	81	&	4804	&	270	&	11.37	&	21.89	&	0.23	&	0.07	\\
UGC 01886	&	-	&	67.4$^2$	&	0.26	&	5	&	57	&	4854	&	502	&	25.67	&	121.29	&	0.85	&	0.33	\\
UGC 01913	&	NGC 0925	&	9.2	&	5.36	&	14	&	58	&	552	&	222	&	3.85	&	12.90	&	0.64	&	<0.01	\\
UGC 01993	&	-	&	107.7$^1$	&	0.89	&	28	&	75	&	8018	&	526	&	13.70	&	95.49	&	1.24	&	1.03	\\
UGC 02045	&	NGC 0972	&	21.7	&	1.66	&	10	&	61	&	1525	&	332	&	2.12	&	45.93	&	4.42	&	<0.01	\\
UGC 02069	&	-	&	36.6$^1$	&	0.62	&	7	&	55	&	3780	&	255	&	4.15	&	17.71	&	1.10	&	0.07	\\
UGC 02080	&	IC 0239	&	10.0	&	2.13	&	6	&	24	&	902	&	135	&	5.46	&	11.52	&	0.16	&	0.01	\\
UGC 02082	&	-	&	14.7	&	2.56	&	11	&	79	&	702	&	215	&	1.36	&	4.64	&	0.04	&	0.01	\\
UGC 02141	&	NGC 1012	&	24.7	&	1.04	&	8	&	60	&	987	&	233	&	2.20	&	17.53	&	1.36	&	1.33	\\
UGC 02154	&	NGC 1023	&	10.2	&	3.71	&	11	&	70	&	695	&	482	&	2.21	&	44.83	&	0.78	&	0.01	\\
UGC 02183	&	NGC 1056	&	21.7	&	0.93	&	6	&	61	&	1540	&	290	&	3.65	&	18.80	&	0.98	&	0.01	\\
UGC 02459	&	-	&	32.4	&	1.17	&	11	&	83	&	2467	&	337	&	12.30	&	31.48	&	0.59	&	1.28	\\
UGC 02487	&	NGC 1167	&	72.2$^3$	&	0.91	&	19	&	41	&	4953	&	468	&	16.65	&	261.23	&	3.21	&	1.36	\\
UGC 02503	&	NGC 1169	&	32.4	&	1.66	&	16	&	54	&	2391	&	461	&	9.69	&	95.99	&	1.12	&	0.12	\\
UGC 02800	&	-	&	18.9$^1$	&	1.17	&	6	&	60	&	1187	&	217	&	2.01	&	5.05	&	1.52	&	0.25	\\
UGC 02855	&	-	&	14.4$^1$	&	1.77	&	7	&	65	&	1196	&	453	&	6.35	&	49.22	&	2.22	&	0.02	\\
UGC 02916	&	-	&	68.0$^2$	&	0.66	&	13	&	24	&	4517	&	336	&	23.12	&	94.12	&	2.45	&	1.40	\\
UGC 03013	&	NGC 1530	&	25.4	&	0.91	&	7	&	55	&	2459	&	341	&	8.98	&	53.03	&	2.07	&	0.03	\\
UGC 03137	&	-	&	22.1	&	1.90	&	12	&	78	&	993	&	216	&	4.41	&	9.32	&	0.15	&	0.02	\\
UGC 03205	&	-	&	47.6$^2$	&	0.66	&	9	&	66	&	3588	&	436	&	9.21	&	65.30	&	0.95	&	2.18	\\
UGC 03326	&	-	&	77.6$^1$	&	1.66	&	37	&	84	&	4060	&	532	&	19.48	&	135.84	&	2.38	&	0.05	\\
UGC 03334	&	NGC1961	&	59.5$^3$	&	2.23	&	39	&	50	&	3935	&	660	&	39.72	&	422.71	&	9.24	&	0.26	\\
UGC 03354	&	-	&	52.5$^1$	&	0.83	&	13	&	70	&	3085	&	441	&	8.89	&	68.85	&	3.22	&	0.02	\\
UGC 03382	&	-	&	67.2$^3$	&	0.63	&	12	&	21	&	4501	&	205	&	5.74	&	73.75	&	0.76	&	0.34	\\
UGC 03407	&	-	&	39.3$^2$	&	0.56	&	6	&	45	&	3602	&	312	&	1.75	&	22.06	&	0.70	&	1.17	\\
UGC 03422	&	-	&	77.2$^2$	&	0.91	&	20	&	62	&	4065	&	416	&	11.05	&	73.40	&	1.08	&	1.00	\\
UGC 03546	&	NGC 2273	&	17.9	&	1.15	&	6	&	53	&	1836	&	339	&	1.95	&	19.09	&	0.56	&	0.21	\\
UGC 03574	&	-	&	17.1	&	0.74	&	4	&	30	&	1441	&	150	&	3.21	&	6.70	&	0.35	&	0.02	\\
UGC 03580	&	-	&	25.9	&	1.07	&	8	&	57	&	1198	&	236	&	3.81	&	12.87	&	0.48	&	0.02	\\
UGC 03642	&	-	&	67.4$^2$	&	0.76	&	15	&	41	&	4498	&	410	&	37.21	&	146.52	&	1.99	&	0.89	\\
UGC 03734	&	NGC 2344	&	23.0	&	1.02	&	7	&	24	&	972	&	150	&	1.12	&	14.80	&	0.11	&	0.01	\\
UGC 03759	&	NGC 2347	&	88.3$^1$	&	0.83	&	21	&	44	&	4416	&	468	&	22.39	&	200.89	&	5.72	&	0.40	\\
UGC 03993	&	-	&	66.3$^3$	&	0.42	&	8	&	24	&	4365	&	175	&	7.13	&	50.87	&	0.91	&	0.04	\\
UGC 04036	&	NGC 2441	&	44.7$^1$	&	1.00	&	13	&	24	&	3469	&	141	&	4.07	&	31.61	&	0.89	&	0.11	\\
UGC 04165	&	NGC 2500	&	15.0	&	1.23	&	5	&	25	&	515	&	113	&	0.97	&	6.82	&	0.35	&	<0.01	\\
UGC 04256	&	NGC 2532	&	51.6$^2$	&	0.83	&	12	&	34	&	5256	&	175	&	6.73	&	56.96	&	3.70	&	0.10	\\
UGC 04273	&	NGC 2543	&	26.3	&	1.23	&	9	&	62	&	2473	&	317	&	4.32	&	20.75	&	1.23	&	0.11	\\
UGC 04284	&	NGC 2541	&	11.2	&	1.51	&	5	&	59	&	559	&	210	&	4.91	&	8.32	&	0.08	&	<0.01	\\
UGC 04458	&	NGC 2599	&	68.6$^3$	&	0.77	&	15	&	32	&	4757	&	285	&	12.52	&	128.09	&	1.39	&	1.09	\\
UGC 04605	&	NGC 2654	&	22.7	&	2.23	&	15	&	78	&	1354	&	430	&	6.32	&	35.50	&	0.82	&	0.01	\\
UGC 04666	&	NGC 2685	&	16.0	&	2.18	&	10	&	58	&	876	&	303	&	1.96	&	17.39	&	0.14	&	0.22	\\
UGC 04806	&	NGC 2770	&	25.5	&	1.73	&	13	&	76	&	1945	&	353	&	5.42	&	19.53	&	0.64	&	0.85	\\
UGC 04838	&	NGC 2776	&	36.0	&	1.07	&	11	&	65	&	2626	&	202	&	6.24	&	44.41	&	1.53	&	0.03	\\
UGC 04862	&	NGC 2782	&	42.1	&	1.62	&	20	&	42	&	2540	&	196	&	4.12	&	67.81	&	4.49	&	0.01	\\
UGC 05060	&	NGC 2893	&	24.0	&	0.51	&	4	&	36	&	1700	&	187	&	0.92	&	6.66	&	0.42	&	0.05	\\
UGC 05079	&	NGC 2903	&	8.5	&	6.01	&	15	&	63	&	555	&	390	&	3.95	&	39.88	&	0.95	&	<0.01	\\
UGC 05251	&	NGC 3003	&	19.6$^1$	&	2.39	&	14	&	77	&	1481	&	294	&	8.89	&	20.10	&	0.40	&	0.02	\\
UGC 05253	&	NGC 2985	&	16.6	&	1.82	&	9	&	36	&	1324	&	316	&	11.62	&	55.21	&	0.82	&	0.37	\\
UGC 05351	&	NGC 3067	&	20.6	&	1.02	&	6	&	71	&	1487	&	281	&	0.91	&	15.67	&	1.17	&	0.01	\\
UGC 05452	&	NGC 3118	&	20.6	&	1.04	&	6	&	78	&	1348	&	216	&	3.41	&	5.91	&	0.07	&	0.02	\\
UGC 05459	&	-	&	25.8	&	1.90	&	14	&	79	&	1108	&	282	&	4.82	&	18.06	&	0.48	&	0.02	\\
UGC 05532	&	NGC 3147	&	39.8	&	2.04	&	24	&	29	&	2812	&	390	&	9.50	&	227.06	&	4.95	&	0.13	\\
UGC 05556	&	NGC 3187	&	26.4	&	1.12	&	9	&	71	&	1582	&	276	&	1.09	&	5.48	&	0.48	&	0.06	\\
UGC 05557	&	NGC 3184	&	13.0	&	3.71	&	14	&	21	&	593	&	146	&	3.95	&	32.47	&	0.20	&	0.01	\\
UGC 05589	&	NGC 3206	&	25.8	&	1.15	&	9	&	59	&	1162	&	182	&	2.61	&	6.50	&	0.03	&	0.06	\\
UGC 05685	&	NGC 3254	&	21.8	&	1.17	&	7	&	72	&	1359	&	378	&	4.71	&	24.34	&	0.22	&	0.10	\\
UGC 05717	&	NGC 3259	&	24.0	&	0.85	&	6	&	58	&	1675	&	242	&	6.34	&	14.71	&	0.43	&	0.05	\\
UGC 05786	&	NGC 3310	&	20.0	&	0.95	&	6	&	40	&	989	&	221	&	3.36	&	17.42	&	3.23	&	0.02	\\
UGC 05789	&	NGC 3319	&	13.3	&	1.82	&	7	&	61	&	739	&	215	&	3.36	&	6.76	&	0.06	&	0.01	\\
UGC 05840	&	NGC 3344	&	10.0	&	3.38	&	10	&	25	&	589	&	175	&	3.01	&	17.40	&	0.25	&	<0.01	\\
UGC 05906	&	NGC 3380	&	26.1	&	0.77	&	6	&	27	&	1600	&	130	&	0.42	&	9.70	&	0.15	&	0.01	\\
UGC 05909	&	NGC 3381	&	25.7	&	1.00	&	7	&	26	&	1633	&	146	&	2.12	&	9.07	&	0.34	&	0.02	\\
UGC 05918	&	-	&	10.0	&	1.23	&	4	&	12	&	338	&	78	&	0.25	&	0.57	&	0.09	&	<0.01	\\
UGC 05997	&	NGC 3403	&	20.2	&	1.38	&	8	&	68	&	1261	&	303	&	4.09	&	12.89	&	0.46	&	0.03	\\
UGC 06024	&	NGC 3448	&	24.0	&	1.48	&	10	&	73	&	1369	&	299	&	6.76	&	21.06	&	1.12	&	0.05	\\
UGC 06128	&	NGC 3512	&	26.1	&	0.79	&	6	&	29	&	1388	&	187	&	0.98	&	13.03	&	0.35	&	0.01	\\
UGC 06225	&	NGC 3556	&	9.6	&	1.99	&	6	&	65	&	698	&	341	&	3.48	&	22.10	&	0.81	&	0.02	\\
UGC 06263	&	NGC 3583	&	33.0	&	1.12	&	11	&	56	&	2134	&	346	&	6.65	&	69.34	&	2.59	&	0.07	\\
UGC 06283	&	NGC 3600	&	14.4	&	0.93	&	4	&	72	&	713	&	218	&	2.86	&	6.14	&	0.26	&	0.01	\\
UGC 06537	&	NGC 3726	&	17.1	&	2.62	&	13	&	47	&	864	&	284	&	5.05	&	35.11	&	0.46	&	0.01	\\
UGC 06621	&	NGC 3786	&	40.0	&	0.97	&	11	&	59	&	2745	&	418	&	4.56	&	42.88	&	1.27	&	0.02	\\
UGC 06778	&	NGC 3893	&	17.1	&	1.35	&	7	&	58	&	968	&	311	&	4.76	&	31.95	&	1.59	&	0.99	\\
UGC 06786	&	NGC 3900	&	22.5	&	1.29	&	8	&	61	&	1801	&	426	&	3.33	&	25.43	&	0.24	&	0.56	\\
UGC 06787	&	NGC 3898	&	22.1	&	1.73	&	11	&	54	&	1170	&	446	&	3.96	&	57.99	&	0.96	&	1.12	\\
UGC 06833	&	NGC 3930	&	12.6	&	1.35	&	5	&	42	&	918	&	161	&	0.99	&	7.00	&	0.36	&	0.01	\\
UGC 06870	&	NGC 3953	&	19.2$^1$	&	3.09	&	17	&	62	&	1051	&	403	&	2.35	&	72.36	&	0.30	&	0.09	\\
UGC 06884	&	NGC 3963	&	49.1$^2$	&	1.26	&	18	&	27	&	3189	&	131	&	8.21	&	68.56	&	1.76	&	0.09	\\
UGC 06930	&	-	&	17.1	&	0.71	&	4	&	42	&	778	&	141	&	2.52	&	4.77	&	0.19	&	0.01	\\
UGC 06964	&	NGC 4010	&	19.1$^1$	&	1.55	&	9	&	78	&	905	&	278	&	1.40	&	8.52	&	0.28	&	0.01	\\
UGC 07030	&	NGC 4051	&	17.2	&	2.45	&	12	&	40	&	704	&	241	&	1.43	&	33.28	&	0.86	&	0.01	\\
UGC 07081	&	NGC 4088	&	14.5$^1$	&	3.54	&	15	&	68	&	756	&	381	&	4.15	&	32.18	&	1.51	&	0.01	\\
UGC 07095	&	NGC 4100	&	20.3$^1$	&	2.29	&	14	&	74	&	1075	&	402	&	3.02	&	35.53	&	1.21	&	0.03	\\
UGC 07183	&	NGC 4157	&	18.0	&	3.08	&	16	&	80	&	771	&	422	&	6.29	&	54.12	&	1.69	&	0.03	\\
UGC 07222	&	NGC 4183	&	16.4$^1$	&	2.13	&	10	&	81	&	931	&	247	&	2.95	&	7.98	&	0.20	&	0.01	\\
UGC 07256	&	NGC 4203	&	15.1	&	1.69	&	7	&	65	&	1088	&	270	&	2.34	&	33.99	&	0.10	&	<0.01	\\
UGC 07321	&	-	&	6.0	&	2.39	&	4	&	86	&	407	&	210	&	0.34	&	0.72	&	0.08	&	0.01	\\
UGC 07399	&	NGC 4288	&	9.2	&	0.85	&	2	&	41	&	535	&	165	&	0.74	&	1.35	&	0.35	&	0.01	\\
UGC 07483	&	NGC 4359	&	16.3	&	0.69	&	3	&	53	&	1271	&	199	&	1.13	&	3.14	&	0.21	&	0.01	\\
UGC 07489	&	NGC 4369	&	11.2	&	1.00	&	3	&	17	&	1029	&	88	&	0.43	&	4.83	&	0.33	&	0.02	\\
UGC 07506	&	NGC 4384	&	36.0	&	0.63	&	7	&	39	&	2532	&	176	&	1.13	&	12.87	&	0.84	&	0.20	\\
UGC 07766	&	NGC 4559	&	8.7	&	5.24	&	13	&	63	&	814	&	256	&	5.43	&	16.21	&	0.22	&	<0.01	\\
UGC 07989	&	NGC 4725	&	12.4	&	4.89	&	18	&	45	&	1210	&	398	&	5.02	&	71.56	&	1.06	&	1.03	\\
UGC 08307	&	NGC5033	&	19.1$^1$	&	9.77	&	54	&	65	&	875	&	425	&	10.43	&	88.98	&	1.76	&	1.27	\\
UGC 08403	&	NGC 5112	&	18.5	&	1.51	&	8	&	52	&	969	&	215	&	3.12	&	8.53	&	0.35	&	0.01	\\
UGC 08699	&	NGC 5289	&	30.9	&	1.17	&	11	&	72	&	2518	&	352	&	2.76	&	18.81	&	0.23	&	0.02	\\
UGC 08709	&	NGC 5297	&	30.9	&	1.86	&	17	&	76	&	2405	&	414	&	12.73	&	50.61	&	1.05	&	0.09	\\
UGC 08711	&	NGC 5301	&	20.2	&	1.99	&	12	&	78	&	1508	&	336	&	3.56	&	18.33	&	0.81	&	0.04	\\
UGC 08863	&	NGC 5377	&	28.0	&	1.82	&	15	&	67	&	1791	&	382	&	2.24	&	47.16	&	0.42	&	0.01	\\
UGC 08900	&	NGC 5395	&	52.7$^2$	&	1.26	&	19	&	62	&	3458	&	565	&	11.21	&	143.67	&	3.53	&	0.25	\\
UGC 09242	&	-	&	27.9	&	2.08	&	17	&	86	&	1438	&	215	&	3.20	&	6.43	&	0.21	&	0.01	\\
UGC 09366	&	NGC 5676	&	38.9	&	1.82	&	21	&	63	&	2121	&	462	&	6.41	&	137.46	&	5.76	&	1.09	\\
UGC 09431	&	NGC 5714	&	38.7$^1$	&	1.41	&	16	&	80	&	2242	&	356	&	7.51	&	29.58	&	0.66	&	0.14	\\
UGC 09644	&	-	&	97.9$^3$	&	0.57	&	16	&	20	&	6664	&	136	&	7.12	&	42.84	&	1.28	&	0.11	\\
UGC 09753	&	NGC 5879	&	15.5	&	1.90	&	9	&	68	&	771	&	287	&	1.32	&	10.88	&	0.28	&	0.01	\\
UGC 09797	&	NGC 5905	&	46.6$^1$	&	1.62	&	22	&	50	&	3393	&	374	&	22.70	&	73.72	&	2.56	&	1.25	\\
UGC 09858	&	-	&	32.2	&	1.95	&	18	&	78	&	2615	&	386	&	10.67	&	28.75	&	0.41	&	1.66	\\
UGC 09969	&	NGC 5985	&	43.7	&	1.99	&	25	&	60	&	2515	&	542	&	10.76	&	144.58	&	1.19	&	0.14	\\
UGC 10359	&	NGC 6140	&	16.0	&	1.04	&	5	&	44	&	908	&	221	&	5.41	&	11.61	&	0.14	&	0.01	\\
UGC 10445	&	-	&	18.1	&	0.95	&	5	&	45	&	962	&	159	&	2.23	&	6.94	&	0.16	&	0.05	\\
UGC 10448	&	NGC 6186	&	154.0$^2$	&	0.79	&	35	&	41	&	11352	&	118	&	9.56	&	439.96	&	8.82	&	0.02	\\
UGC 10470	&	NGC 6217	&	23.0	&	1.12	&	7	&	34	&	1355	&	192	&	5.94	&	30.66	&	1.86	&	0.02	\\
UGC 10497	&	-	&	65.6$^2$	&	0.59	&	11	&	65	&	4296	&	267	&	8.93	&	21.34	&	0.36	&	0.17	\\
UGC 10564	&	NGC 6237	&	21.0	&	0.62	&	4	&	52	&	1129	&	175	&	5.64	&	11.08	&	0.27	&	0.03	\\
UGC 11124	&	-	&	25.0	&	1.12	&	8	&	26	&	1599	&	153	&	2.23	&	11.71	&	0.16	&	0.03	\\
UGC 11218	&	NGC 6643	&	20.6	&	1.66	&	10	&	61	&	1484	&	350	&	3.20	&	30.27	&	1.78	&	0.04	\\
UGC 11269	&	NGC 6667	&	44.9	&	0.93	&	12	&	56	&	2581	&	412	&	13.36	&	66.90	&	1.73	&	0.02	\\
UGC 11283	&	IC 1291	&	30.0	&	0.66	&	6	&	35	&	1946	&	198	&	2.55	&	9.44	&	0.38	&	0.55	\\
UGC 11429	&	NGC 6792	&	62.2$^1$	&	1.04	&	19	&	58	&	4637	&	510	&	12.26	&	129.81	&	1.60	&	0.31	\\
UGC 11466	&	-	&	18.1	&	0.74	&	4	&	53	&	821	&	237	&	2.79	&	10.96	&	0.84	&	0.01	\\
UGC 11670	&	NGC 7013	&	15.0	&	2.08	&	9	&	71	&	775	&	340	&	1.35	&	26.16	&	0.28	&	<0.01	\\
UGC 11852	&	-	&	82.4$^2$	&	0.46	&	11	&	44	&	5845	&	328	&	26.73	&	82.67	&	0.96	&	0.45	\\
UGC 11861	&	-	&	14.4	&	0.89	&	4	&	61	&	1482	&	259	&	2.10	&	10.25	&	0.47	&	0.02	\\
UGC 11909	&	-	&	14.1	&	1.00	&	4	&	78	&	1105	&	242	&	2.87	&	7.78	&	0.39	&	0.01	\\
UGC 11914	&	NGC 7217	&	15.0	&	2.29	&	10	&	35	&	950	&	301	&	0.70	&	52.88	&	0.68	&	<0.01	\\
UGC 11951	&	NGC 7231	&	14.2	&	0.85	&	4	&	69	&	1086	&	223	&	1.56	&	4.97	&	0.35	&	0.92	\\
UGC 11994	&	-	&	65.8$^1$	&	1.04	&	20	&	82	&	4882	&	436	&	6.95	&	57.44	&	1.69	&	0.29	\\
UGC 12554	&	NGC 7640	&	9.9	&	4.06	&	12	&	78	&	363	&	238	&	3.05	&	8.78	&	0.24	&	<0.01	\\
UGC 12693	&	-	&	60.5$^1$	&	0.55	&	10	&	78	&	4958	&	236	&	9.67	&	15.50	&	0.69	&	0.12	\\
UGC 12732	&	-	&	15.1	&	1.38	&	6	&	28	&	728	&	131	&	1.96	&	4.06	&	0.59	&	0.01	\\
UGC 12754	&	NGC 7741	&	13.6$^1$	&	1.82	&	7	&	49	&	749	&	202	&	1.78	&	5.76	&	0.36	&	0.01	\\
UGC 12808	&	NGC 7769	&	61.5$^3$	&	0.87	&	16	&	68	&	4225	&	326	&	4.79	&	134.32	&	6.21	&	0.05	\\

\noalign{\vspace{2pt}}\hline
\noalign{\vspace{5pt}}

\multicolumn{10}{l}{$^1$ Distance from Cosmicflows-2 catalogue.} \\
\multicolumn{10}{l}{$^2$ Distance from NED catalogue.} \\
\multicolumn{10}{l}{$^3$ Distance from Hubble flow with Virgo infall corrected systemic velocity.} \\
\multicolumn{10}{l}{$^4$ Baryonic mass from baryonic Tully-Fisher relation.} \\
\end{longtable}

\end{document}